\documentclass[amsmath, amssymb, aps, prx, reprint, superscriptaddress]{revtex4-2}

\usepackage{hyperref}

\usepackage{siunitx}
\usepackage{amsmath}
\usepackage{nicefrac}
\usepackage{tabularx}
\usepackage{makecell}
\usepackage{multirow}

\usepackage{varwidth}
\usepackage{amssymb}

\usepackage{booktabs}
\usepackage{enumitem}

\usepackage{float}

\usepackage{graphicx}
\graphicspath{{./figures}}

\usepackage{algpseudocode}

\usepackage{algorithm}

\def\rec{\mathcal{R}}

\def\mem{\kern-1pt\triangleright\kern-1pt}
\def\den{\rho}
\def\sup{\textsc{s}}

\DeclareSIUnit\pixel{px}
\makeatletter

\newcommand{\algcom}[2][.45]{\hfill\Comment{\begin{minipage}[t]{#1\linewidth} \begin{flushleft} #2 \end{flushleft}\end{minipage}}}

\newcommand{\iter}{\jmath}

\algdef{SE}[REPEATN]{Repeatn}{EndRepeatn}[1]{\algorithmicrepeat\ #1 \textbf{times}}{\algorithmicend$\,$ \algorithmicrepeat}

\newcommand{\affila}{Laboratory for Solid State Physics, ETH Zurich, 8093 Zürich, Switzerland}
\newcommand{\affilb}{Paul Scherrer Institut, 5232 Villigen, Switzerland}
\newcommand{\affilc}{Max Planck Institute for the Structure and Dynamics of Matter, 22761 Hamburg, Germany}
\newcommand{\affild}{University of Oulu, 90570 Oulu, Finland}
\newcommand{\affile}{European XFEL, 22869 Schenefeld, Germany}
\newcommand{\affilf}{Ludwig Maximilian University Munich, 85748, Garching, Germany}
\newcommand{\affilg}{Max Planck Institute of Quantum Optics, 85748 Garching, Germany}
\newcommand{\affilh}{University of Rostock, 18059 Rostock, Germany}
\newcommand{\affili}{Technische Universitat Berlin, Institut fur Optik und Atomare Physik, 10623 Berlin, Germany}
\newcommand{\affilj}{Department of Cell and Molecular Biology, Uppsala University, 75124 Uppsala, Sweden}
\newcommand{\affilk}{Department for the Ecology of Animal Societies, Max Planck Institute of Animal Behavior, 78467 Konstanz, Germany}
\newcommand{\affill}{Karlsruhe Institute of Technology (KIT), Institute of Nanotechnology, 76131 Karlsruhe, Germany}
\newcommand{\affilm}{Department of Applied Physics, AlbaNova University Center, KTH Royal Institute of Technology, S-106 91 Stockholm, Sweden}
\newcommand{\affiln}{Department of Physics and Astronomy, Uppsala University, 753 10 Uppsala, Sweden}
\newcommand{\affilo}{MAX IV, 224 84 Lund, Sweden}
\newcommand{\affilp}{Laboratory for Ultrafast X-ray Sciences, Institute of Chemical Sciences and Engineering, Ecole Polytechnique Federale de Lausanne (EPFL), Lausanne, Switzerland}
\newcommand{\affilq}{Physics Department, Università degli Studi di Milano, 20133 Milano, Italy}

\begin{document}                  
\title{SPRING: an effective and reliable framework for image reconstruction in single-particle Coherent Diffraction Imaging}

\author{Alessandro \surname{Colombo}}\email{alcolombo@phys.ethz.ch}\affiliation{\affila}
\author{Mario \surname{Sauppe}}\affiliation{\affila}
\author{Andre \surname{Al Haddad}}\affiliation{\affilb}
\author{Kartik \surname{Ayyer}}\affiliation{\affilc}
\author{Morsal \surname{Babayan}}\affiliation{\affild}
\author{Rebecca \surname{Boll}}\affiliation{\affile}
\author{Ritika \surname{Dagar}}\affiliation{\affilf}\affiliation{\affilg}
\author{Simon \surname{Dold}}\affiliation{\affile}
\author{Thomas \surname{Fennel}}\affiliation{\affilh}
\author{Linos \surname{Hecht}}\affiliation{\affila}
\author{Gregor \surname{Knopp}}\affiliation{\affilb}
\author{Katharina \surname{Kolatzki}}\affiliation{\affila}
\author{Bruno \surname{Langbehn}}\affiliation{\affili}
\author{Filipe R. N. C. \surname{Maia}}\affiliation{\affilj}
\author{Abhishek \surname{Mall}}\affiliation{\affilc}
\author{Parichita \surname{Mazumder}}\affiliation{\affilc}
\author{Tommaso \surname{Mazza}}\affiliation{\affile}
\author{Yevheniy \surname{Ovcharenko}}\affiliation{\affile}
\author{Ihsan Caner \surname{Polat}}\affiliation{\affila}
\author{Dirk \surname{Raiser}}\affiliation{\affile}
\author{Julian C. \surname{Schäfer-Zimmermann}}\affiliation{\affila}\affiliation{\affilk}
\author{Kirsten \surname{Schnorr}}\affiliation{\affilb}
\author{Marie Louise \surname{Schubert}}\affiliation{\affila}\affiliation{\affill}
\author{Arezu \surname{Sehati}}\affiliation{\affilm}\affiliation{\affilj}
\author{Jonas A. \surname{Sellberg}}\affiliation{\affilm}
\author{Björn \surname{Senfftleben}}\affiliation{\affile}
\author{Zhou \surname{Shen}}\affiliation{\affilc}
\author{Zhibin \surname{Sun}}\affiliation{\affilb}
\author{Pamela H. W. \surname{Svensson}}\affiliation{\affiln}
\author{Paul \surname{Tümmler}}\affiliation{\affilh}
\author{Sergey \surname{Usenko}}\affiliation{\affile}
\author{Carl Frederic \surname{Ussling}}\affiliation{\affila}
\author{Onni \surname{Veteläinen}}\affiliation{\affild}
\author{Simon \surname{Wächter}}\affiliation{\affila}
\author{Noelle \surname{Walsh}}\affiliation{\affilo}
\author{Alex V. \surname{Weitnauer}}\affiliation{\affila}
\author{Tong \surname{You}}\affiliation{\affilj}
\author{Maha \surname{Zuod}}\affiliation{\affila}
\author{Michael \surname{Meyer}}\affiliation{\affile}
\author{Christoph \surname{Bostedt}}\affiliation{\affilb}\affiliation{\affilp}
\author{Davide E. \surname{Galli}}\affiliation{\affilq}
\author{Minna \surname{Patanen}}\affiliation{\affild}
\author{Daniela \surname{Rupp}}\affiliation{\affila}

\begin{abstract}
Coherent Diffraction Imaging (CDI) is an experimental technique to gain images of isolated structures by recording the light scattered off the sample. In principle, the sample density can be recovered from the scattered light field through a straightforward Fourier Transform operation. However, only the amplitude of the field is recorded, while the phase is lost during the measurement process and has to be retrieved by means of suitable, well-established phase retrieval algorithms. In this work, we present SPRING, an analysis framework tailored to X-ray Free Electron Laser (XFEL) single-shot single-particle diffraction data that implements the Memetic Phase Retrieval method to mitigate the shortcomings of conventional algorithms. We benchmark the approach on experimental data acquired in two experimental campaigns at SwissFEL and European XFEL. Imaging results on isolated nanostructures reveal unprecedented stability and resilience of the algorithm's behavior on the input parameters, as well as the capability of identifying the solution in conditions hardly treatable so far with conventional methods. A user-friendly implementation of SPRING is released as open-source software, aiming at being a reference tool for the coherent diffraction imaging community at XFEL and synchrotron facilities.
\end{abstract}

\maketitle

\section{Introduction}

The advent of Free-Electron Lasers (FELs) \cite{schneider2010flash, emma2010first, allaria2012highly, ishikawa2012compact, decking2020mhz, kang2017hard, milne2017swissfel} represented a game-changer for the study of matter and its interaction with light. Short-wavelength FELs are capable of producing extremely bright pulses of coherent and monochromatic light, ranging from the extreme ultraviolet (XUV) spectrum to hard X-rays. A unique feature of those pulses is their time duration, often on the order of a few tens of femtoseconds and nowadays approaching the attosecond time scale \cite{duris2020tunable, yan2024terawatt, franz2024terawatt}. Such exceptionally short pulses, in combination with their high brightness, unlocked unprecedented opportunities for time-resolved investigations of ultrafast processes of molecules \cite{dold2023melting} and even electrons \cite{chapman2019x,rupp2020imaging,nishiyama2019ultrafast}, as well as structural studies on especially fragile systems \cite{feinberg2022x,niozu2020characterizing,niozu2021crystallization,colombo2023imaging}.
Several experimental techniques can be exploited at XFEL (XUV and X-ray FEL) facilities, gaining immense benefits from the properties of their light pulses. This is particularly true for imaging methods, which intrinsically require a large amount of photons.

Imaging techniques aim at retrieving structural properties of the sample, and, specifically, the spatial distribution of their electron density, to which photons are sensitive in this energy range. Due to the intensity of the light and the high photon energy direct imaging methods like microscopy are hardly viable due to the lack of suitable optics. For this reason, imaging at XFELs is performed in an \emph{indirect} manner, by recording the coherent light scattered off the sample and then retrieving its spatial distribution by sophisticated data analysis approaches. This \emph{lensless} experimental scheme is called Coherent Diffraction Imaging (CDI) \cite{marchesini2003x, seibert2011single, hantke2014high, van2015imaging, ekeberg2024observation, colombo2023three}.

While various approaches can be linked to the definition of CDI, like Bragg-CDI \cite{pfeifer2006three} or Ptychography \cite{rodenburg2007hard}, the most common technique employed at XFELs is Single Particle CDI (SP-CDI) in the small-angle scattering regime \cite{chapman2010coherent, miao2011coherent, seibert2011single}. Here, the light scattered by an isolated sample is acquired by a detector, which is placed sufficiently far from the interaction region to satisfy the \emph{far-field} condition \cite{chapman2006high}. The diffraction signal is collected for a scattering angle up to a few degrees, such that the portion of the Ewald's sphere defined by the corresponding momentum transfer accessible by the detector can be well approximated with a flat surface. Under Born's approximation, the field scattered at small angles is then proportional to the Fourier Transform (FT) of the sample's electronic density projected along the axis of the XFEL beam. 

Despite the straightforward mathematical relationship, which in principle enables the restoration of the sample's 2D projection by an inverse FT operation on the scattered field, the latter cannot be fully accessed by photon detectors. Detectors are, in fact, two-dimensional arrays of photon counters. The acquired signal is then only sensitive to the intensity of the diffracted field, while its phase is completely lost in the measurement process \cite{kirian2020imaging, fannjiang2020numerics, nugent2010coherent, thibault2010x}.

The field's phases are however retrievable if the sample is confined in space and ``small enough'', i.e. if the so-called \emph{oversampling condition} is met \cite{kirian2020imaging, colombo2023imaging}. In such a case, it can be mathematically shown that the information contained in the field's intensity alone is sufficiently redundant to also carry information on its phases \cite{grohs2020phase}. This possibility is exploited by phase retrieval algorithms \cite{marchesini2007invited, shechtman2015phase, fannjiang2020numerics} to successfully retrieve the lost phases and, consequently, the corresponding image of the sample's electronic density.

The core idea behind \emph{phase retrieval algorithms} was originally conceived in the 1980' by Fienup \cite{fienup1982phase} and then further developed in the following years \cite{marchesini2007invited}. These algorithms are based on the iterative application of two constraints. Given a starting guess for the sample's electronic density, they go back and forth from the Fourier representation to the real representation of the spatial density. In the reciprocal space, the signal's amplitudes are replaced by those experimentally measured by the detector. In the real space, the oversampling condition is enforced by constraining the limited spatial extension of the electronic density. With proceeding iterations, the similarity between the FT amplitude of the retrieved density and the experimental data is optimized.

A known issue of phase retrieval algorithms is their tendency to stagnate in local optima and/or being unstable \cite{marchesini2007invited}.
On top of that, XFEL diffraction data present additional challenges. First, the central scattering region is not recorded, due to a hole in the detector to avoid damages from the FEL transmitted beam. Second, the brightness of the patterns, i.e. the total number of photons recorded by the detector, fluctuates on a shot-by-shot basis, with little or no possibility to be tuned. Third, samples often have unpredictable shapes, giving little or no a-priori information to facilitate the reconstruction process. 
As a result, the identification of the correct reconstruction by phase retrieval algorithms is not ensured, it is possible only in a restricted range of conditions and it requires a significant case-by-case tuning of the algorithm parameters by an experienced person. 

These issues and limitations are well known by the CDI community, and a significant effort has been invested into developing more reliable algorithmic methods. The use of several phase retrieval processes with different starting conditions on the same diffraction data has proved to be particularly effective.
On the one hand, the ensemble of final reconstruction outcomes can be compared and statistically analyzed to identify those reconstruction procedures that reached the correct solution and exclude the failed attempts \cite{van2015imaging,sekiguchi2016classification,favre2020free,takayama2024similarity}. On the other hand, the simultaneous execution of several reconstruction processes with different starting conditions allows for more sophisticated optimization strategies based on the sharing of information between the ensemble of ongoing reconstructions \cite{chen2007application,yoshida2024protocol,takayama2024similarity}.

Recently, we developed a sophisticated approach to the \emph{phase retrieval problem} in CDI that conceptually belongs to this ensemble-based class. The method is called Memetic Phase Retrieval (MPR)\cite{colombo2017facing, colombo2018high}. It is based on a natural computing scheme and it had been originally designed to deal with Electron Diffraction Imaging data \cite{abbey2008keyhole,de2012keyhole,carlino2018coherent}. In this work, we present the SPRING framework, which implements an improved version of the MPR method tailored on the specific features of single-particle single-shot experiments at FELs. Together with this publication, we release its software implementation as an open-source Python3 module, named \emph{spring}, conceived to exploit the full capability of multi-CPU and multi-GPU computing systems. The pre-release of the SPRING software is currently being used by different research groups recently involved in the analysis of CDI experiments. The \emph{spring} module, its documentation and code can be accessed at \url{https://share.phys.ethz.ch/~nux/software/spring/main/docs/}.

This article is a comprehensive report of SPRING, its implementation of the MPR method and its performance on experimental XFEL diffraction data. 
After a brief introduction to the \emph{phase retrieval problem} in Sec. \ref{sec:phaseproblem}, Sec. \ref{sec:MPR} is dedicated to an intuitive description of the method. 
In the following Sec. \ref{sec:expdata} we challenge the algorithm on experimental data. Its performance is compared with conventional approaches, and its behavior, capabilities and limitations on images at the limits of the capabilities of standard \emph{phase retrieval} algorithms are probed. The experimental diffraction patterns used as a benchmark have been acquired with the two main CDI detector types of today's XFEL imaging instruments: the pnCCD detector\cite{kuster20211} of the SQS instrument at the European XFEL\cite{mazza2023beam,tschentscher2017photon} and the Jungfrau detector\cite{mozzanica2018jungfrau,hinger2022advancing} installed at the Maloja endstation of SwissFEL\cite{sun2022ultrafast}. 
Further details of the SPRING implementation are reported and discussed in Appendices \ref{sec:memeticops} and \ref{subsec:missint}, aided by an extensive use of pseudo-code.

The quality and reliability of the imaging results, the superior performance and the highly optimized and handy software implementation, make SPRING a strong candidate for becoming a reference analysis tool, fulfilling the currently high demand for a reliable and easy-to-use analysis software tailored on single-particle single-shot CDI data at XFELs.

\section{The phase problem and conventional approaches} \label{sec:phaseproblem}

The \emph{phase problem} arises when a function $\den(\vec{x})$ has to be retrieved, but only its Fourier modulus $M(\vec{q}) = \left| \mathcal{F} \left[ \den \right] (\vec{q}) \right|$ is known.
In single-shot CDI experiments, under the assumptions of \emph{small-angle} and \emph{far-field} scattering, and validity of \emph{Born's approximation}, the recorded diffraction intensities $I(\vec{q})$ acquired by a detector correspond to $M^2(\vec{q})$, while $\rho(\vec{x})$ is the electronic density of the sample under study \cite{kirian2020imaging} projected along the beam direction. 
The phases of the scattered field, lost in the measurement process, have to be retrieved by solving a two-dimensional phase retrieval problem.

The phase problem is known to have a unique solution if the density $\rho$ that produces the scattering signal is limited in space, i.e. it is of limited size and surrounded by an environment of known scattering density (usually vacuum, with value $0$). This condition is enforced by a so-called support function $S(\vec{x})$, which assumes values equal to $1$ only where the scattering density is known to have non-zero values, or $0$ otherwise \cite{fienup1978reconstruction,marchesini2003x}. The spatial extension of the support function and, thus, of the object's density, has to be sufficiently restricted to satisfy the so-called \emph{oversampling condition} \cite{kirian2020imaging}. If such a condition is met, it can be mathematically demonstrated that the phase retrieval problem has a unique solution \cite{grohs2020phase} (up to some straightforward ambiguities, see \cite{colombo2023imaging}), i.e. there exists only one density $\rho$ fully ``contained'' in the support function and whose Fourier amplitude corresponds to the measured diffraction data.

Iterative phase retrieval algorithms \cite{fienup1978reconstruction,marchesini2007invited} cyclically apply the \emph{support constraint} and the \emph{amplitude constraint} to a starting density $\den$. The  \emph{support constraint} is enforced by suppressing the density values in real-space that are located outside the support function. The \emph{amplitude constraint} is enforced by replacing the Fourier modulus with the experimentally recorded one, $M(\vec{q})$.

The error $E[\den ]$ of a density $\den$ is defined as the difference between the Fourier amplitude of the density $\left| \mathcal{F} \left[ \den \right]\right|$   and the experimental data $M$. The scope of a reconstruction process is to minimize the error $E$ (see Appendix \ref{subsec:optprob} for a more formal description).

The simplest phase retrieval algorithm is the Error Reduction (ER) \cite{fienup1978reconstruction}, which takes its name from its characteristics of \emph{always} reducing the error value iteration after iteration. This feature makes the ER algorithm very \emph{stable}, but also prone to \emph{stagnation} in local minima of the error function $E$. 

To circumvent this limitation, more advanced iterative phase retrieval algorithms have been conceived \cite{marchesini2007invited,shechtman2015phase}, like the Hybrid Input-Output (HIO) \cite{fienup1982phase} and the  Relaxed Averaged Alternating Reflections (RAAR) \cite{luke2004relaxed}, which make a more sophisticated use of the support and amplitude constraints. The great advantage of those algorithms is that they suffer less from stagnation in local optima, thus being more \emph{ergodic}, i.e. they can better explore the parameter space. However, they are often unstable, meaning that they fail to stably converge towards an optimum of the error function. From now on, we will refer to those iterative algorithms as IAs.

As a trade-off, a common approach is to alternate a number $\iter$ of \emph{ergodic} IAs and a number $\iter_\text{ER}$ of ER algorithm \cite{marchesini2007invited}. This means that a sequence of algorithms is used, and can be formalized as:
\begin{equation}\label{eq:sequence}
\den' \gets \textsc{Sequence} \left(\den \right)
\end{equation}
At the beginning of the reconstruction process, when the density guess $\den$ is far from the solution, a high number of IAs iterations $\iter$ is set to better explore the space. This number is reduced during the reconstruction down to $\iter=0$ when the reconstruction is supposed to be close to the solution, to reduce instabilities and let the process converge to an optimum via the ER algorithm.

The support function $S(\vec{x})$ defines the ``shape'' of the sample density, and it is necessary to ensure the existence of a unique solution. 
In most experiments, and in particular single-shot single-particle CDI, the spatial extension of the sample is not known a-priori, which in turn means that the support $S$ cannot be correctly defined before the reconstruction process. 

This aspect was an open problem until the Shrink-wrap (SW) algorithm \cite{marchesini2003x}, which improves the shape of the support function $S$ along with the reconstruction process of the density $\rho$. This operation is performed by first smoothing the density with a gaussian kernel of standard deviation $\sigma$, and then  defining the new support function $S'$ as those coordinates where the values of the smoothed density $\rho_\sigma$ are greater than a fraction $\tau$  of the maximum value of $\rho_\sigma$. 
It is possible to indicate the SW algorithm as a function \textsc{UpdateSupport}. The \textsc{UpdateSupport} function calculates a new support function $S'$ for the input density $\rho$, and depends on the parameters $\sigma$ and $\tau$:
\begin{equation}\label{eq:shrinkwrap}
S' \gets \textsc{UpdateSupport}\left( \rho \right)
\end{equation}
The execution of the SW algorithm for updating the support function is typically alternated with the execution of an algorithm sequence in Eq. \eqref{eq:sequence} \cite{marchesini2003x}.

As a reconstruction process involves not only the sample density $\rho$, but also its support $S$, it is convenient to indicate a reconstruction as $\rec$, with its spatial density $\rec\mem\rho$ and its support $\rec\mem\sup$. Typically, a number $P \gg 1$ of reconstruction processes are executed with different random initializations, forming a set of reconstructions $\left\{ \rec^{p} \right\}$. 

\begin{algorithm}[H]
\caption{}
\label{alg:randomsearch}
\begin{algorithmic}[1] 
\State $\left\{ \rec^{p} \right\} \gets $ \Call{Initialize}{} 
\For{$g=1,...,G$}
	\State $\left\{ \rec^{p} \right\} \gets \Call{Improve}{\left\{ \rec^{p} \right\}}$
\EndFor
\State $\rho_\text{out} \gets$ \Call{Best}{$\left\{ \rec^{p}\mem\rho \right\}$}
\end{algorithmic}
\end{algorithm}

A full phase retrieval procedure is summarized in Alg. \ref{alg:randomsearch}.
The \textsc{Initialize} function takes care of randomly initializing a set of $P$ reconstructions $\left\{ \rec^{p} \right\}$, each with its density $\rec^p \mem\rho$ and support $\rec^p \mem\sup$. Then, the reconstructions are improved $G$ times by the function \textsc{Improve}, defined in Alg. \ref{alg:normalimprove}. It internally executes a sequence of iterative algorithms (see Eq. \eqref{eq:sequence}) and the support update function (see Eq. \eqref{eq:shrinkwrap}) on all the $P$ reconstructions in the set $\left\{ \rec^{p} \right\}$.
At the end of the process, the best results in $\left\{ \rec^{p} \right\}$ are kept as final reconstruction output.

\begin{algorithm}[H]
\caption{}
\label{alg:normalimprove}
\begin{algorithmic}[1] 
\Function{Improve}{$\left\{ \rec^{p} \right\}$}		
	\For{$p=1,...,P$}	
		\State $\rec^p\mem\rho \gets \Call{Sequence}{\rec^p\mem\den}$
		\State $\rec^p\mem\sup \gets \Call{UpdateSupport}{ \rec^p\mem\den }$	
	\EndFor
	\State \Return $\left\{ \rec^{p} \right\}$
\EndFunction 
\end{algorithmic}
\end{algorithm}

The Memetic Phase Retrieval (MPR) approach \cite{colombo2017facing} implemented in SPRING reshapes this typical workflow to make a more efficient use of the information provided by the several reconstruction processes, as discussed in the next section.

\section{Memetic Phase Retrieval} \label{sec:MPR}

\begin{figure*}
  \centering
  \includegraphics[width=0.99\linewidth]{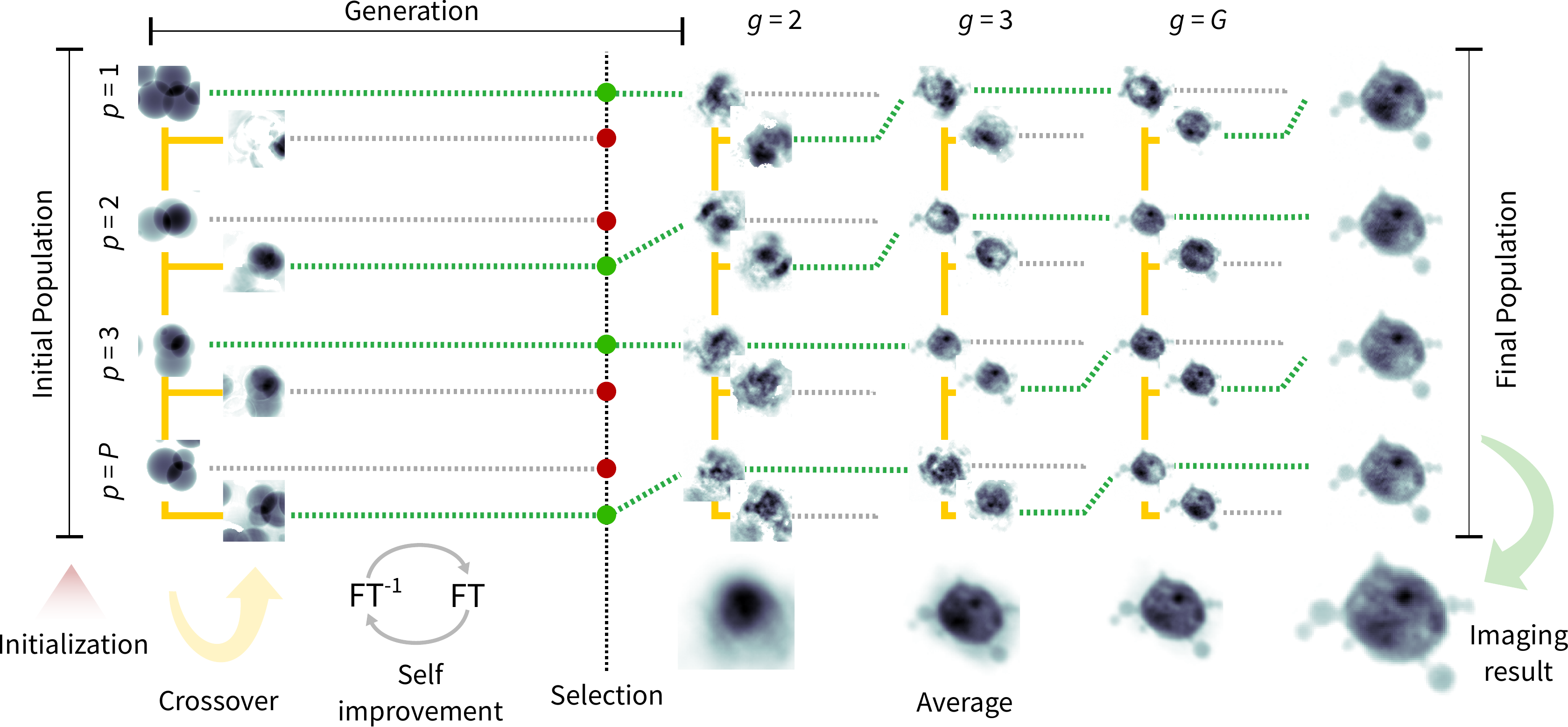}
  \caption{Scheme of the MPR method implemented into SPRING. On the left of the figure, a population of reconstructions of size $P$ is initialized. For each \emph{generation} $g$ of the algorithm, three operations are performed. First, a new population of the same size is created by combining density values from individuals of the original population via the \emph{crossover} operator. In the following \emph{self-improvement} step the reconstructions of both populations are locally optimized by alternating sequences of iterative phase retrieval algorithms, to update the density values, and the \emph{Shrink-wrap} algorithm, to update the corresponding support functions. At the end of each generation, reconstructions of the two populations are compared for each index $p$, and only those that reached the lowest error value survive to the next generation. With the proceeding generations $g$, the population of reconstructions converges towards a single solution. The convergence can be identified by observing the average reconstruction for each generation, shown in the bottom. The average density at the final generation $g=G$ is kept as reconstruction result of the whole imaging process.}
  \label{fig:flowchart}
\end{figure*}

The structure of the Memetic Phase Retrieval (MPR) is inspired by Memetic Algorithms, which merge together local optimization methods with Genetic Algorithms \cite{moscato2002memetic}. In fact, MPR alternates the execution of conventional iterative phase retrieval algorithms with additional operations, presented in this section, to better minimize the error of the CDI reconstruction.
The overall workflow of the MPR approach implemented in SPRING retains its original main structure \cite{colombo2017facing}. It is reported in Alg. \ref{alg:mpr} and accompanied by a qualitative but intuitive sketch in Fig. \ref{fig:flowchart}. 
The MPR method is based on the simultaneous execution of $P$ reconstructions, identified by the set $\left\{ \rec^{p} \right\}$, on which the action of conventional iterative phase retrieval algorithms is intertwined with operations inherited from Genetic Algorithms.

\begin{algorithm}[H]
\caption{The MPR main structure}
\label{alg:mpr}

\begin{algorithmic}[1]
\State $\left\{ \rec^{p} \right\} \gets $ \Call{Initialize}{}\label{l:mpr:init}

\For{$g=1,...,G$}\label{l:mpr:genloop}

	\State $ \left\{ \rec_\text{new}^{p} \right\}  \gets \Call{Crossover}{\left\{ \rec^{p} \right\}}$ \label{l:mpr:crossover} 

	\State $\left\{ \rec^{p} \right\}  \gets$ \Call{Improve}{$\left\{ \rec^{p} \right\}$} \label{l:mpr:improve} 
	\State $\left\{ \rec_\text{new}^{p} \right\}  \gets$ \Call{Improve}{$\left\{ \rec_\text{new}^{p} \right\}$} \label{l:mpr:improvenew}

	\State $\left\{ \rec^{p} \right\} \gets$ \Call{Select}{$\left\{ \rec^{p} \right\}, \left\{ \rec_\text{new}^{p} \right\}$}  \label{l:mpr:select} 

\EndFor
\State $\rho_\text{out} \gets$ \Call{Average}{$\left\{ \rec^{p}\mem\rho \right\}$} \label{l:mpr:mean}
\end{algorithmic}
\end{algorithm}

The first step of the algorithm (leftmost side of Fig. \ref{fig:flowchart}) consists of the creation of a set of initial guesses for the density. This operation is performed by the \textsc{Initialize} function (line \ref{l:mpr:init} of Algorithm \ref{alg:mpr}), which takes care of filling with starting random values the density matrices $\rec^{p}\mem\den$  and the respective support matrices $\rec^{p}\mem\sup$ for each reconstruction $\rec^{p}$, similarly to the standard approach in Algorithm \ref{alg:randomsearch}. 
The set $\left\{ \rec^{p} \right\}$ will be referred to as a \emph{population}, borrowing the terminology from Evolutionary Algorithms (EAs). This initialization step is not strictly connected to the MPR method, and its implementation, discussed in Appendix \ref{subsec:init}, has been specifically designed for SPRING.

After the initialization step, the main body of the algorithm is composed by the loop at line \ref{l:mpr:genloop} of Algorithm \ref{alg:mpr}. We will refer to each iteration of such a loop as \emph{generation}, again borrowing the term from EAs. For each \emph{generation}, the core functions \textsc{Crossover}, \textsc{Improve} and \textsc{Select} are called. 

The \textsc{Crossover} routine creates a new population of $P$ reconstructions  $ \left\{ \rec_\text{new}^{p} \right\}$ by combining the information provided by the current population $\left\{ \rec^{p} \right\}$ , as sketched in Fig. \ref{fig:flowchart}. Each new reconstruction $\rec_\text{new}^{p}$ is obtained by mixing the density and support values of four different reconstructions in the original population $\left\{ \rec^{p} \right\}$. Details on the crossover implementation in SPRING are given in Appendix \ref{subsec:crossover}.
How information is mixed with the crossover operation has been significantly improved over the original MPR method \cite{colombo2017facing}, to better deal with intrinsic correlation in the diffraction data (see the tiling operation in Appendix \ref{subsec:crossover}) and with the missing intensities in the center (see Appendix \ref{subsubsec:crossavg}).

The \textsc{Improve} function is analogue to the one defined for the conventional approach in Alg. \ref{alg:normalimprove} and improves the density estimates via iterative algorithms and the support functions via the Shrink-wrap algorithm. Here, the operation is performed on the original population $\left\{ \rec^{p} \right\}$ as well as on the new population $ \left\{ \rec_\text{new}^{p} \right\}$. 
SPRING implements an enhanced version of iterative algorithms that allows for a mitigation of high-resolution artifacts produced by missing data in the gaps between the scattering detector tiles (see Sec. \ref{subsec:intbound}) and a more refined execution flow of iterative phase retrieval algorithms with respect to the original MPR implementation.
Additional information on this step, known as \emph{self-improvement} in the context of Memetic Algorithms (and indicated as such in Fig. \ref{fig:flowchart}), is given in Appendix \ref{subsec:improve}.

The \textsc{Select} procedure, whose details are discussed in Appendix \ref{subsec:selection}, compares each individual of the original population, $\rec^{p}$, with the corresponding reconstruction in the new one $\rec_\text{new}^{p}$ right after the \emph{self-improvement} step. Only those reconstructions that performed ``better'' than their counterpart (i.e. reached a lower error value) survive to the next generation while others are discarded, as schematically reported in Fig. \ref{fig:flowchart}.

The \emph{crossover}, \emph{self-improvement} and \emph{selection} operations are repeated for a given number of generations $G$, during which the population of reconstructions converges towards the solution. The progress of the process can be monitored by calculating the average of the reconstructions in the population at each generation $g$, reported on the lower side of Fig. \ref{fig:flowchart}. The average reconstruction is a valuable indicator of the performance of the retrieval process, because it presents only the spatial features of the sample that are common to most of the retrieved densities in the population. For the same reason, the average density calculated at the last generation is kept as the final output of the MPR algorithm, i.e. the solution to the imaging problem.

The detailed description and discussion on how the three operations \textsc{Crossover}, \textsc{Improve} and \textsc{Select} are implemented in SPRING, along with details on how the starting population is initialized, are deferred to Appendix \ref{sec:memeticops}.
The following part of the manuscript focuses on characterizing and understanding the behavior of SPRING on experimental data acquired at XFEL facilities, and comparing its output and performance with those provided by a conventional use of iterative phase retrieval algorithms.

\section{Dealing with experimental data} \label{sec:expdata}

The data treated in this work were generated from two CDI experiments conducted at SwissFEL \cite{prat2020compact} and European XFEL \cite{decking2020mhz}. 
Sample diffraction patterns, produced by single XFEL pulses, and corresponding CDI reconstructions performed with SPRING are reported in Fig. \ref{fig:dataexamples}.
Technical details on the datasets are given below, while further information on the data selection is provided in Sec. 1 of the Supplemental Material.

\begin{figure*}[!]
  \centering
  \includegraphics[width=1.\linewidth]{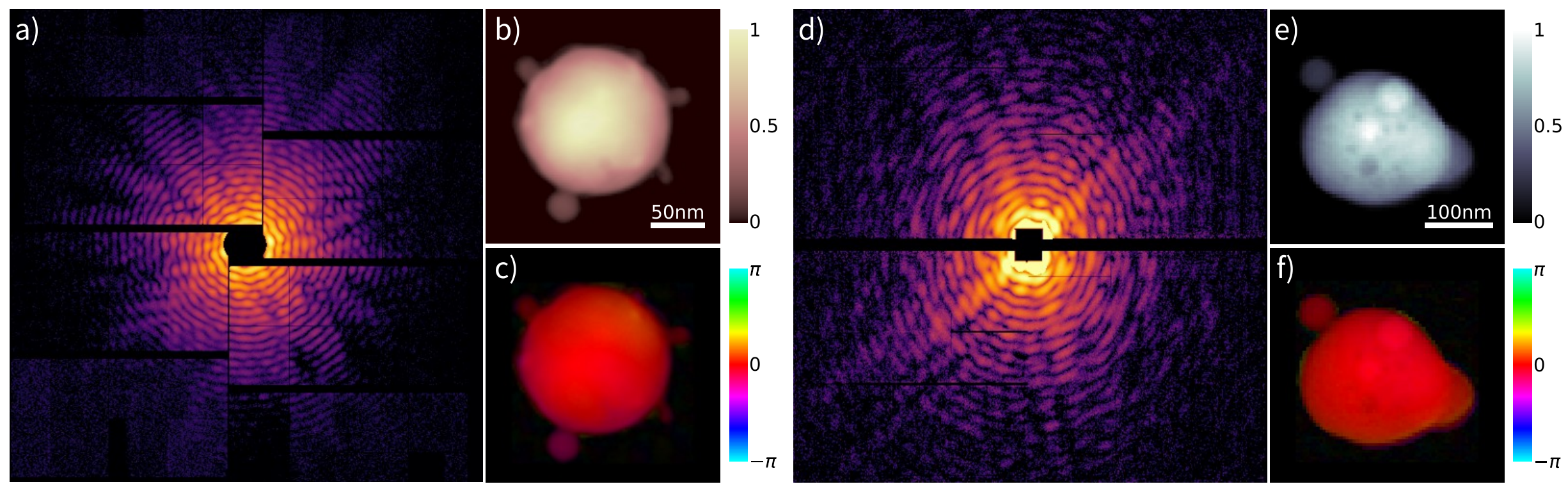}
  \caption{Examples of experimental diffraction patterns. In a), a diffraction pattern extracted from dataset A. The corresponding reconstruction, performed with SPRING, is reported in b) for the amplitude, and c) for the phases. In d), an example extracted from dataset B is shown, with reconstructed amplitude and phase reported in e) and f).}
  \label{fig:dataexamples}
\end{figure*}

\vspace{1em}

\textbf{Dataset A} -- Pure argon and mixed argon/xenon nanoclusters produced by a pulsed cluster source equipped with an Even-Lavie valve \cite{pentlehner2009rapidly} are imaged at \SI{700}{\electronvolt} and \SI{1}{\kilo\electronvolt} photon energy, equivalent to \SI{1.77}{\nano\meter} and \SI{1.24}{\nano\meter} wavelength, respectively. 
The XFEL delivered single pulses of \SI{50}{\femto\second} duration and \SI{2.5}{\milli\joule} total energy, focused on a sub-\SI{3}{\micro\meter} spot.
 The experiment was performed at the Maloja endstation of SwissFEL \cite{sun2022ultrafast}. The beamline is equipped with a 4 Mpixel Jungfrau detector optimized for soft x-ray photon detection \cite{mozzanica2018jungfrau,hinger2022advancing}. The detector pixels, with physical size of $\SI{75}{\micro\meter} \times \SI{75}{\micro\meter}$, are organized in separated tiles of $256 \times 512$ pixel and arranged for the detection of light up to \ang{13.5} scattering angle. An example of a scattering pattern belonging to this dataset is shown in Fig. \ref{fig:dataexamples}a.

\vspace{1em}

\textbf{Dataset B} -- Nanoparticles produced from an aerosolised aqueous solution of calcium chloride and sodium chloride mixture are dried and injected into the experimental chamber in form of aerosol, and are imaged at around \SI{1.1}{\kilo \electronvolt} photon energy, equivalent to \SI{1.13}{\nano\meter} wavelength. A constant output atomiser (model 3076, TSI Inc., MI, USA) was used to create a stream of polydisperse aerosol focused to the interaction region using an aerodynamic lens. 
The experiment was performed at the Small Quantum Systems (SQS) instrument of the European XFEL \cite{mazza2023beam,tschentscher2017photon} using the Nano-sized Quantum Systems (NQS) end station. Typically, the XFEL was providing soft X-ray pulses with pulse energies of up to \SI{5}{\milli\joule} and the beam was focused to a diameter of about \SI{5}{\micro\meter} in the interaction region.
The scattering light is recorded by a \SI{1}{Mpixel} pnCCD detector \cite{kuster20211}. The detector is composed of two modules of $512 \times 1024$ pixel, with size $\SI{75}{\micro\meter} \times \SI{75}{\micro\meter}$ each, resulting in a total linear dimension of \SI{76.8}{mm}. Its placement allows the scattered light to be recorded up to \ang{6.3} at the edge of the detector. An example of a scattering pattern taken from this dataset is shown in Fig. \ref{fig:dataexamples}d.

\vspace{1em}

The \emph{phase retrieval problem} is known to be particularly challenging \cite{chen2019gradient,sun2018geometric} and there are different aspects in which the standard use of iterative phase retrieval algorithms present some shortcomings and limitations \cite{shechtman2015phase, thibault2006reconstruction, marchesini2007invited}. Among those, the following are of particular relevance when dealing with experimental data:

\begin{itemize}[leftmargin=*]
\item The convergence of a reconstruction to the correct solution is not ensured. It is furthermore challenging to identify a correct reconstruction in an automatized manner.
\item The algorithm's performance is dependent on the settings and the initialization of the densities. Among those, the ones that define the starting support size and how the support is upgraded during the process are of particular relevance for single-shot XFEL data, where thousands of samples of unknown size and shape are imaged with different brightness and noise.
\item Scattering information is missing in some regions of the detector, in particular in the central part due to the presence of a hole to avoid damage from the transmitted XFEL beam (see the patterns in Fig. \ref{fig:dataexamples}). This makes algorithms particularly unstable, it renders reconstructions more prone to artifacts for increasing sample size and it limits in practice the field of view of the CDI method.
\end{itemize}

After a general description of SPRING's behavior in a typical reconstruction procedure, the following subsections challenge the SPRING framework on these aspects and compare its performance to conventional approaches.

\subsection{Monitoring the reconstruction process} \label{subsec:monitor}

\begin{figure*}
  \centering
  \includegraphics[width=0.99\linewidth]{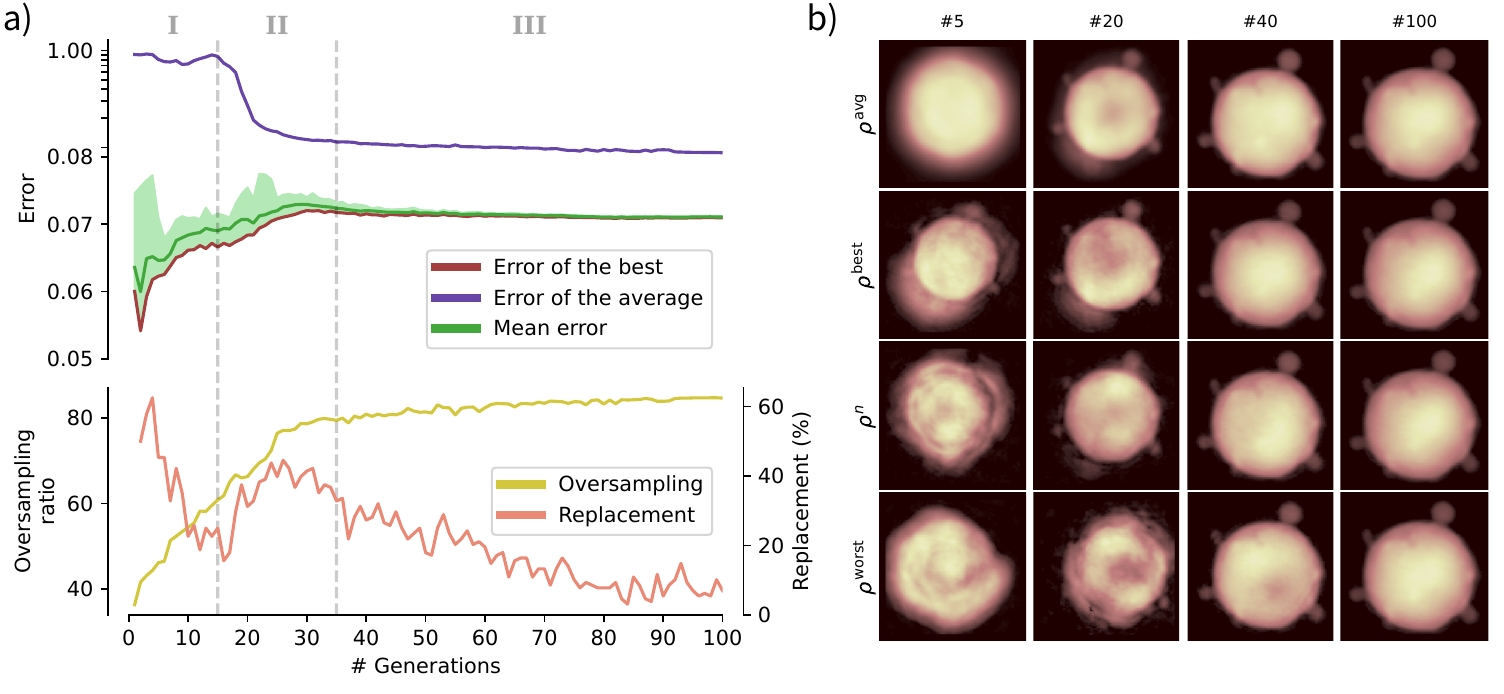}
  \caption{Behavior of a typical reconstruction procedure with SPRING on a diffraction pattern extracted from Dataset A. In a), the value of different performance indicators are reported as a function of the MPR generations. Their characteristics allow the identification of three distinct stages of the reconstruction process, I, II and III, highlighted by the gray dashed lines. A direct insight into the population of reconstructions at generations 5, 20, 40 and 100 is given in b). The first row reports the average reconstruction, indicated as $\den^\text{avg}$, while the second row shows the reconstruction with the lowest error $\den^\text{best}$. The third row is the reconstruction at the $50^\text{th}$ percentile of the errors distribution, while the last row corresponds to the reconstruction with the highest error $\den^\text{worst}$. Please refer to the main text for details.}
  \label{fig:monitor}
\end{figure*}

This section is dedicated to an overall description of a typical reconstruction process with SPRING, summarized in Fig. \ref{fig:monitor}.
The behavior of these indicators as a function of the algorithm generations for a typical phase retrieval routine are shown in Fig. \ref{fig:monitor}a. Fig. \ref{fig:monitor}b reports, instead, insights into the population of reconstructions at the different steps. The reconstruction has been performed by using the HIO algorithm in the \emph{self-improvement} step of MPR (see Appendix \ref{subsec:improve}), with a starting number of iterations $\iter=20$ (see Appendix \ref{subsubsec:sequence}) and a starting support size $\Sigma^\text{init} = \SI{60}{\text{pixel}}$ (see Appendix \ref{subsec:init}). The other parameters are kept to their default value indicated in Table \ref{tab:params}.

The upper side of Fig. \ref{fig:monitor}a reports the error value. For a given reconstruction $\rec$, its value is calculated as it follows:
\begin{equation}\label{eq:error_calc}
E[\rec] = \sqrt{\frac{\sum_{ij} \Xi_{ij} \cdot \left( \left| \mathcal{F}\left[ \rec\mem\sup \cdot \rec\mem\den \right]_{ij} \right| - M_{ij}  \right)^2  }{\sum_{ij} \Xi_{ij} \cdot M_{ij}^2 }}
\end{equation}
where $\Xi_{ij}$ is the mask function, whose value is $\Xi_{ij}=1$ where the intensity values of $M_{ij}$ are known, and $\Xi_{ij}=0$ otherwise. The operation $\rec\mem\sup \cdot \rec\mem\den$ operation is the element-wise multiplication between the support matrix $\rec\mem\sup_{ij}$ and the density matrix $\rec\mem\den_{ij}$. The result of this operation is a density perfectly compatible with the \emph{support constraint}, i.e. with zero values outside the support function (see also Appendix \ref{subsec:optprob} for further discussion).

Eq. \eqref{eq:error_calc} calculates the difference between the retrieved Fourier amplitude and the experimental one, $M_{ij} = \sqrt{I_{ij}}$. It is worth noting that this value is normalized with the L$^2$ norm of the experimental modulus $M_{ij}$. On the one hand, this does not influence the reconstruction process, as this normalization factor is a constant value for a given diffraction pattern. On the other hand, this normalization allows to obtain error values that can be interpreted as \emph{relative deviations} and that have similar order of magnitude (often between $10^{-1}$ and $10^{-2}$) across different experimental patterns.

Different error values are reported in Fig. \ref{fig:monitor}a. The red line represents the error of the best reconstruction $\rec^\text{best}$, i.e. the one with the lowest error value. The region indicated in light green covers the region between the best reconstruction and the worst one, i.e. the one with highest error value. The line marked in dark green is, instead, the mean value of the error across the whole population. A further error value is reported in blue, corresponding to that of the average reconstruction. The average reconstruction $\rec^\text{avg}$ is computed as it follows:
\begin{equation}\label{eq:avgrec}
\begin{split}
\rec^\text{avg} \mem \rho_{ij} & = \frac{\sum_{p=0}^P \rec^{p} \mem \rho_{ij}}{P} \\
\rec^\text{avg} \mem \sup_{ij} & = \frac{\bigvee_{p=0}^P \rec^{p} \mem \sup_{ij}}{P}
\end{split}
\end{equation}
where the operation $\bigvee_{p=0}^P$ indicates the logical \textsf{OR} operation. In practice, the support function $\rec^\text{avg} \mem \sup$ of the average reconstruction is given by the union of all supports of the population of reconstructions  $\left\{ \rec^{p}\right\}$.

The lower plot in Fig. \ref{fig:monitor}a reports two additional quantities. One is termed \emph{oversampling}, and gives information about the size of the support function of the best reconstruction. It is calculated via the formula:
\begin{equation}\label{eq:oversampling}
O_d = \frac{N_p^2}{\sum_{ij} \rec^\text{best} \mem \sup_{ij}}
\end{equation}
which expresses the ratio between the total number of entries in the matrix (of shape $N_p \times N_p$) and the actual extension of the support function. For example, $O_d=60$ means that the extension of the support function is a $60^{\text{th}}$ of the overall matrix. Higher values of $O_d$ correspond to smaller supports, and vice versa. 

The last quantity shown in Fig. \ref{fig:monitor}a is the \emph{replacement factor}, and gives indications about the effectiveness of the selection step (see Appendix \ref{subsec:selection}). For each generation, the \emph{selection} operation compares each individual of the original population of candidate solutions  $\rec_{p}$ with the corresponding one in the new population $\rec^{p}_\text{new}$, with the latter replacing the former if it reached a lower error value. The replacement factor reports how many new reconstructions have performed better than the corresponding ones in the old population and replaced them. A value of \SI{0}{\percent} means that no replacement happened, i.e. the new reconstructions $\rec^{p}_\text{new}$ performed systematically worse than the parents $\rec^{p}$. Instead, a value of \SI{100}{\percent} implies that all the new reconstructions performed better and replaced the original ones. 

From the behavior of the aforementioned performance indicators it is convenient to identify three different stages, highlighted as I, II and III in the upper part of Fig. \ref{fig:monitor}a. 

\subsubsection*{Stage I: Shape identification} In this first part of the reconstruction procedure (up to generation 15 in the example reported in \ref{fig:monitor}a) the algorithm tries to identify the correct spatial extension of the sample. Typically, the starting support function is larger than the final correct one, and its size is reduced via the action of the SW algorithm, as indicated by the increasing value of the oversampling ratio reported as a yellow line in Fig. \ref{fig:monitor}a. The shrinking of the support is strictly correlated to an increase in the reconstructions errors, highlighted by the red and green lines in Fig. \ref{fig:monitor}a. This behavior highlights well how the size of the support function influences the error value, and clarifies why a comparison of the error values is only meaningful between reconstructions with the same support size (see Appendix \ref{subsubsec:modshrink}). In this phase, when the correct sample shape is not yet identified, the error of the average reconstruction (blue line in Fig. \ref{fig:monitor}a) is close to 1 and constant, as the average is calculated on widely differing spatial distributions. This aspect is directly visible in the first column of Fig. \ref{fig:monitor}b, which reports a snapshot of the reconstructions at generation 5. There, the best reconstruction (second row) is already outlining a density distribution resembling the correct one, while the rest of the population still have very differing shapes. As a consequence, the average reconstruction in the first row has no clear features. In this first stage, the replacement factor (orange line in Fig. \ref{fig:monitor}a) drops from a high starting value of around \SI{50}{\percent}. At the beginning, the reconstructions are so far away from the solution that the two populations $\rec^{p}$ and $\rec^{p}_\text{new}$ are substantially equivalent, carrying to equal probability of surviving to the next generation. As the support is shrinking it becomes less likely that $\rec^{p}_\text{new}$ replaces $\rec^{p}$. The condition where the replacement factor is below \SI{50}{\percent} but significantly above \SI{0}{\percent} indicates that the \emph{genetic operators} are effective.

\subsubsection*{Stage II: Shape definition} This second phase of the reconstruction process (from generation 15 to generation 35 in Fig. \ref{fig:monitor}a) is characterized by the convergence of the population towards the correct sample shape. While the support function is still shrinking (see the oversampling behavior in Fig. \ref{fig:monitor}a) and, consequently, the error values of the reconstructions are still growing (see the error of the best) similarly to Stage I, there are clear signatures of a different behavior. First, the mean error is getting closer to the error of the best and the overall error distribution (light green area in Fig. \ref{fig:monitor}a)) is shrinking. This is accompanied by a rapid increase in the replacement factor. However, the clearest signature that identifies this stage comes from the error of the average (blue line in Fig. \ref{fig:monitor}a), characterized by a sudden drop. These are signs that the algorithm has identified the correct sample shape and this information is quickly propagating across the population via the crossover and selection operators. It is possible to directly recognize this dynamics in the second column of Fig. \ref{fig:monitor}b, which reports the status of the population at generation number 20. Here, most of the reconstructions are resembling the correct sample shape (second and third row) and, as a consequence, the same happens to the average reconstruction (first row).
  
\subsubsection*{Stage III: Density refinement} Once the sample shape (i.e. the support function) has been identified, the algorithm focuses on the correct retrieval of the density values (from generation 35 in the example reported in Fig. \ref{fig:monitor}a). Here, the oversampling is still slowly rising, thanks to both a further refinement of the support function and the decrease of the $\sigma$ parameter of the SW algorithm over the generations. However, in this phase, the reduction of the support size is correlated to a slow decrease of the error value (differently from the previous stages I and II) both for the individual reconstructions and for the average. This behavior is due to both the action of the crossover and selection operations that promote the reconstructions with lower error, as well as the decreasing number of IAs iterations (HIO) and the corresponding increase of ER iterations (which naturally tends to reach lower errors). The dynamics of the reconstructions population is recognizable in Fig. \ref{fig:monitor}b, which gives an overview on the reconstructions at generation 40 (third column) and 100 (last generation, fourth column), corresponding to the beginning and the end of Stage III, respectively.

\subsection{Comparison with the conventional approach} \label{subsec:comp}

\begin{figure*}
  \centering
  \includegraphics[width=\linewidth]{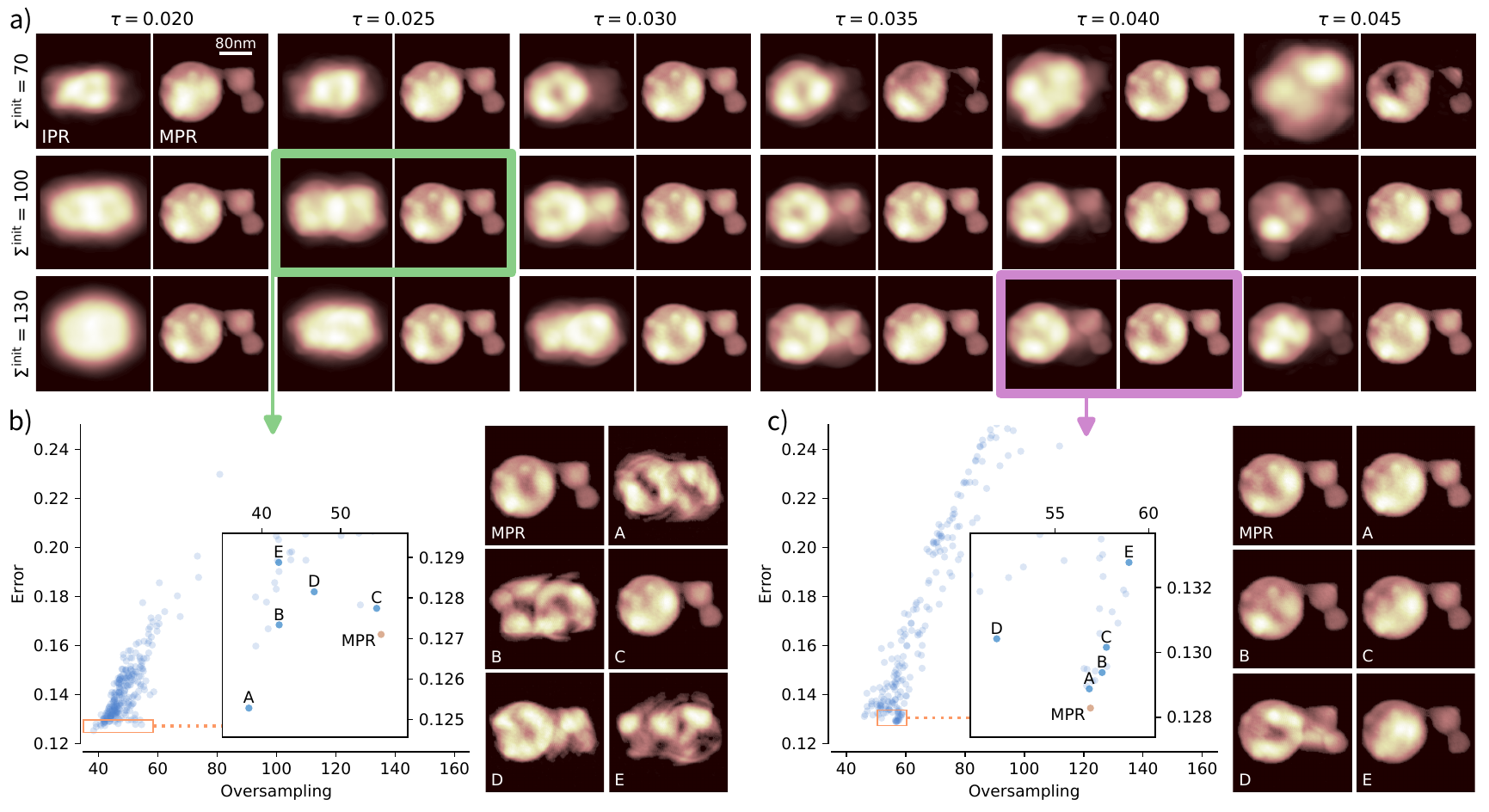}
  \caption{Comparison between the conventional use of iterative algorithms, here indicated as IPR, and the SPRING implementation of the MPR method. The diffraction data belongs to Dataset A. In a), the average reconstruction at the end of the retrieval process is shown for both IPR and MPR, as a function of the threshold value $\tau$ for the support update via the Shrink-wrap algorithm and the size of the starting support function $\Sigma^\text{init}$. Individual reconstructions are inspected in detail for the cases $(\tau,\Sigma^\text{init})=(0.025,100)$ and $(0.04,130)$ in sub-figures b) and c), respectively. The plots in b) and c) report the oversampling degree (on the $x$ axis, inversely proportional to the retrieved support size) and the error value ($y$ axis) for all $256$ individual reconstructions obtained with the IPR method. The inset graph shows a zoom of the main plot in the region around the IPR reconstructions with the lowest error value. In addition, the red dot indicates the error and oversampling value for the best reconstruction in the MPR population. On the right side of the plots in b) and c), six different reconstructions are shown. The best of the MPR population is compared with the best of the IPR population (marked as A) and other four reconstructions (B to E) in the vicinity of A.}
  \label{fig:comparison}
\end{figure*}

The convergence properties of standard algorithms, i.e. their ability to identify the solution, have a relevant dependence on the input parameters. Those that involve the support function, like its initial size and the threshold value $\tau$ for the SW algorithm, are often problematic and strongly influence the convergence to a solution.

This aspect is often a problem when dealing with experimental data because, even within the same dataset, samples typically have different sizes and diffraction patterns are characterized by different brightness (which, in turn, implies a different amount of noise). Human input is frequently required to adjust these parameters almost on an shot-to-shot basis.
 
Fig. \ref{fig:comparison} reports a comparison between the outcome of a conventional use of iterative algorithms (IAs) and the MPR method implemented in SPRING. In this section we refer to the conventional approach with the term IPR. 
The conventional IPR approach is obtained by ``switching off'' the \emph{crossover} and \emph{selection} operations of the MPR framework. In this way, each reconstruction is completely independent and only undergoes the alternated execution of iterative algorithms and the support update with the standard SW algorithm \cite{marchesini2003x}. The population size is set $P=128$ for MPR, while it is doubled to $P=256$ for the conventional IPR to make the comparison more fair in terms of computational cost and time to solution. In fact, MPR executes standard IAs on two sets of reconstructions, the original $\left\{ \rec^{p} \right\}$ and the new $\left\{ \rec_\text{new}^{p} \right\}$ created by the \emph{crossover} operation, doubling in practice the number of optimized reconstructions. The initial number of iterations of IAs is set to $80$, alternated with $20$ iterations of Error Reduction algorithm.

Both IPR and MPR are tested against different values of the initial support size $\Sigma^\textrm{init}$ (expressed in pixels, see Appendix \ref{subsec:algparams}) and the SW algorithm threshold $\tau$ \cite{marchesini2003x} for its update. The retrieved densities reported in Fig. \ref{fig:comparison}a are calculated by averaging the densities of the whole population $\left\{ \rec^p \right\}$. While the average is a meaningful reconstruction output for MPR, this is not the case for the IPR, for which only a fraction of the reconstruction processes actually converge to the right solution. Therefore, a visual evaluation of the IPR average is a qualitative indicator on how many reconstructions in the population have reached convergence.

Fig. \ref{fig:comparison}a shows how MPR can reliably retrieve the sample density across most of the parameters range, with incorrect reconstructions only appearing for higher $\tau$ values ($0.035$ and $0.045$) and $\Sigma^\textrm{init}=70$ (i.e. a starting support size \SI{30}{\percent} lower than the actual maximum extension of the sample of around \SI{100}{pixel}). This example well underlines the resilience of MPR with diverse parameter settings.

While no convergence is reached for the conventional IPR at the lowest threshold value $\tau=0.02$, the correct sample shape can be glimpsed at $\tau=0.025$ for $\Sigma^\textrm{init}=100$. The IPR convergence improves for higher threshold values, up to an optimum between $\tau=0.03$ and $\tau=0.035$ for $\Sigma^\textrm{init}=100$, and $\tau=0.035$ to $\tau=0.04$ for $\Sigma^\textrm{init}=130$. IPR doesn't appear to reach a significant amount of converged reconstructions for the smallest starting support size $\Sigma^\textrm{init}=70$ at any $\tau$ value.

Fig. \ref{fig:comparison}b is a detailed inspection of the individual IPR outcomes for the parameter combinations  $(\tau,\Sigma^\text{init})=(0.025,100)$, i.e. the earliest occurrence of sub-optimal values where IPR reaches the correct solution for some individual reconstructions.
Here, the final error of the reconstructions obtained with the IPR method (see Eq. \eqref{eq:error_calc}) is reported against the oversampling degree given by the reconstructed support function (see Eq. \eqref{eq:oversampling}). Each blue point corresponds to a different reconstruction $\rec^{p}$ in the population of size $P=256$. The overall distribution of the points clearly shows the strong correlation between the size of the reconstructed support and the error value. The inset plot is a zoom of the low-error region. Some points are marked with letters A-E, and the corresponding retrieved densities are reported on the right side. 
The reconstruction marked as ``A'' correspond to the one with the lowest error, while the others are chosen in its neighborhood. Additionally, the best result in the MPR population is reported as a red dot in the inset graph, with the corresponding density shown on the right side of the sub-figure.
A first important consideration concerns the best IPR reconstruction ``A''. Despite having reached the lowest error, the corresponding density is clearly far off from the correct solution. This is caused by a larger retrieved support (i.e. lower oversampling value), and highlights why, in MPR, reconstruction errors are only compared by constraining the same oversampling value for the whole reconstruction population. See Appendix \ref{subsubsec:modshrink} for further details and considerations.
Furthermore, it appears that the best density is the one marked as ``C'', unsurprisingly the closest to the best MPR density (red dot) but only the fifth in terms of error ranking. This well underlines why, in general, identifying reconstructions that have reached convergence in an automatized manner is extremely challenging.
Reconstruction ``C'' and the best of MPR have very similar oversampling degree (i.e., their error can be compared) and the error of the MPR best is lower. This is an indication that, once that the correct oversampling value is identified, MPR can reach a deeper minimum of the error landscape.
The statistical evaluation of the final outcomes as function of the oversampling degree and the error value presented in Fig. \ref{fig:comparison}b is provided to give an intuitive understanding. A less intuitive but more refined metric for the identification of the converged reconstructions is proposed and discussed in Ref. \cite{favre2020free}.

A similar consideration applies to Fig. \ref{fig:comparison}c, where the same overview is reported instead for $\tau$ and $\Sigma^\text{init}$ in the optimal configuration for the conventional IPR method, corresponding to $0.04$ and $130$. The higher number of converged reconstructions for the conventional IPR method is evident by visually inspecting the the densities reported on the right side of Fig. \ref{fig:comparison}c. Differently from Fig. \ref{fig:comparison}b, the best IPR reconstruction here corresponds to the one with the lowest error value (``A''), as the correct oversampling value has been identified. All of the other reconstructions right above ``A'' in the inset graph, i.e. those with lower error values and similar oversampling, have reached a proper convergence. Still, some outliers with low error but too low oversampling appear, as exemplified by the reconstruction ``D''. Again, the best MPR reconstruction, reported as a red dot in the plot, has a lower error value than ``A'', meaning that a better optimization is done by MPR when the correct support is retrieved.
Despite the optimal $\tau$ and $\Sigma^\text{init}$ for the standard IPR have been used, still only around \SI{10}{\percent} of the $256$ reconstructions have reached a neighborhood of the correct solution to the phase problem, and their automatized identification remains challenging. Further tests on the stability of SPRING with different algorithm parameters are given in Sec. 2 of the Supplemental Material.

\begin{figure*}
  \centering
  \includegraphics[width=0.9\linewidth]{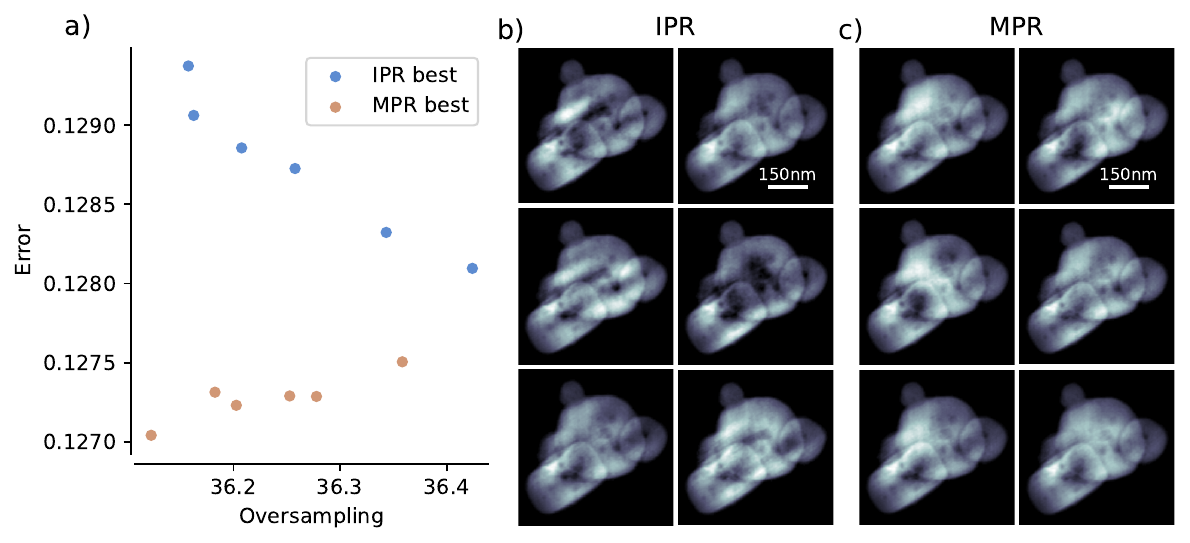}
  \caption{Dependence of the reconstruction result on the density initialization. Six reconstruction procedures, with population size $P=256$ for the conventional IPR and $P=128$ for the SPRING implementation of MPR, are performed on a diffraction pattern selected from dataset B. The random starting densities are created by setting a different \emph{seed} for the random number generator for each of the six cases, to make the procedures statistically independent between each others. The error and oversampling value reached by the best reconstruction in the population for each of the six attempts is reported in a), for both the conventional IPR (blue dots) and MPR (red dots). The corresponding densities are shown in b) for IPR and c) for MPR.}
  \label{fig:seed}
\end{figure*}

In both of the reconstruction attempts investigated in Fig. \ref{fig:comparison}b and Fig. \ref{fig:comparison}c the error of the best MPR density is lower than the one of the best reconstruction for IPR. 
This property is analyzed in more details in Fig. \ref{fig:seed}. Here, six full reconstruction procedures, each involving a population of size $P=128$ reconstructions for MPR and size $P=256$ for IPR, have been performed. 
In SPRING, densities are initialized by filling the real-space matrix with a number of spherical density profiles, randomly defined in size and position, as described in Appendix \ref{subsec:init}. In this test, a different \emph{seed} of the random number generator has been set for each attempt to create different, statistically independent, sets of starting conditions for each reconstruction attempt.

 The algorithms parameters have been optimized to ensure the convergence of the conventional IPR method, corresponding to an initial support size $\Sigma^\text{init}=130$, a threshold value for the support update $\tau=0.035$ and an initial number of IAs iterations equal to $80$ (alternated with $20$ iterations of ER algorithm, see also Appendix \ref{subsec:algparams}).

Fig. \ref{fig:seed}a reports the error value and oversampling reached by the best reconstruction in the population $\left\{ \rec^{p} \right\}$ for each of the six reconstruction attempts. Blue dots indicate the values for the conventional IPR, while red dots correspond to the MPR outcomes. The independent MPR results appear to be systematically better than the ones of the conventional IPR, i.e. the best MPR reconstruction systematically reaches a lower error than the best IPR reconstruction.

The six best densities from the independent reconstruction processes are visually reported in Fig. \ref{fig:seed}b for the conventional IPR and in Fig. \ref{fig:seed}c for the SPRING implementation of MPR. It can be observed that the higher values of the IPR errors are connected with a higher variability in the reconstructed densities in Fig. \ref{fig:seed}b. The MPR results in Fig. \ref{fig:seed}c are, instead, much more similar despite being statistically independent reconstructions. This observation is a further sign that the MPR method is capable of reaching a spatial density closer to the \emph{global optimum} than the conventional IPR approach.

The tests presented in Fig. \ref{fig:comparison} and \ref{fig:seed} are not exhaustive and cannot be considered a thorough evaluation of the performances of the two approaches, the conventional IPR and the SPRING implementation of MPR. Smaller samples with higher oversampling degree, or brighter diffraction patterns, or different number of IAs iterations may shrink or widen the performance gap between the two methods. On the other hand, no fine-tuning of the MPR parameters have been performed for this test, and all parameters not explicitly mentioned in this section have been kept to their default value (see Appendix \ref{subsec:algparams}). We encourage the reader to perform further, more specific, tests by directly using the open-source \emph{spring} Python package.

\subsection{Imaging in challenging conditions} \label{subsec:initparam}

\subsubsection{Extended samples} \label{subsec:large}

\begin{figure*}
  \centering
  \includegraphics[width=.9\linewidth]{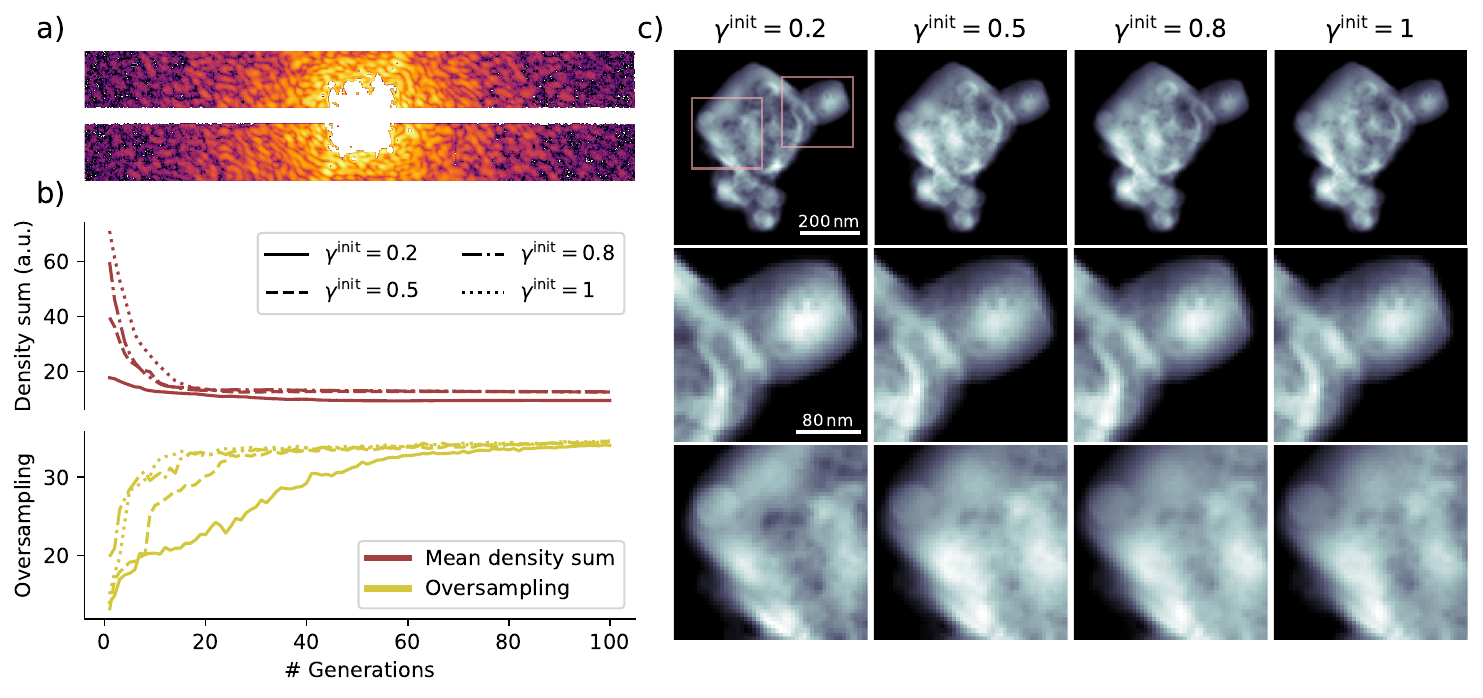}
  \caption{Effects of the gamma value of the starting guesses $\gamma^\text{init}$ on the phase retrieval outcome. This test is performed on a diffraction image from Dataset B. In a), a cutout of the diffraction data is reported. Missing intensities are now present also in those regions where the detector's pixels reached saturation. In b), the sum of the reconstructed density and the oversampling are reported as a function of the generation for different values of $\gamma^\text{init}$. The reconstruction outcomes are compared in c), where the full reconstructed densities are shown in the first row and two closeups of different regions are reported in the second and third rows.}
  \label{fig:initexp}
\end{figure*}

The large spatial extension of the sample, corresponding to particularly low oversampling values $O_d$ (see Eq. \eqref{eq:oversampling}) renders the phase retrieval problem more challenging as the number of unknowns to be retrieved (i.e. the pixel values contained in the support function) is higher. This is particularly true when scattering signal is missing in the center of the pattern (as visible in Fig. \ref{fig:dataexamples}), which is an unavoidable feature of XFEL data.

The lack of data in the center, corresponding to low momentum transfer, causes the arising of \emph{loosely constrained modes} \cite{thibault2006reconstruction}, i.e. density fluctuations in the reconstructions at low resolution. Those fluctuations have a larger and larger impact on the reconstruction quality with increasing sample size, up to the point where, for particularly low oversampling degrees, the reconstruction attempts are unsuccessful. 
The SPRING implementation of MPR includes a particular strategy to deal with these \emph{loosely constrained modes}, whose details are given in Appendix \ref{subsubsec:crossavg}.

Among those modes, the Gaussian mode is typically the most prominent (see Ref. \cite{thibault2006reconstruction}), and its strongest effect is the uncertainty that it introduces in the overall density normalization. Thus, it becomes more challenging to correctly retrieve for samples with larger spatial extension and the success of a reconstruction depends more strongly on the initial density normalization. This fact limits in practice the field of view of the CDI technique. Here we show how the MPR method is resilient to different normalization values for the starting densities.

SPRING creates starting guesses by filling the space with randomly distributed spheres. The density profile of the spheres can be tuned with a parameter $\gamma^\text{init}$, which acts similarly to the \emph{gamma value} in image processing: the closer $\gamma^\text{init}$ to $0$, the flatter the density profiles. The starting densities are then normalized by comparing their L$^2$ norm in Fourier space with the experimental data, calculated only where intensities are experimentally recorded. Different values for $\gamma^\text{init}$ thus have a strong influence on the initial normalization of the densities. See Appendix \ref{subsec:init} for a detailed description of the densities initialization implemented in SPRING.

The implications of different values of $\gamma^\text{init}$ on the final outcome of the SPRING reconstructions are reported in Fig. \ref{fig:initexp}.
A cutout taken from the middle of the diffraction pattern treated in this example is shown in Fig. \ref{fig:initexp}a. This pattern shows strong scattering, with photons recorded up to the edge of the image. Due to the limited dynamic range of the pnCCD detector, there is a saturated region towards the center. 
The dependence of the initial densities normalization on the $\gamma^\text{init}$ value can be recognized in the initial value (generation number $0$) of the mean density sum reported as a red line in Fig. \ref{fig:initexp}b. This initial value ranges from $18$ for the lowest value $\gamma^\text{init}=0.2$ to $70$ for the highest value $\gamma^\text{init}=1$. 

The outcomes of the reconstruction procedures are reported in Fig. \ref{fig:initexp}c. High-resolution details are coherently reconstructed among the three cases, as highlighted by the enlarged details in the second and third row.
However, the overall reconstructions at the first row reveal differences in the low-resolution brightness profile. While these differences are barely perceivable for  $\gamma^\text{init}$ equal to $0.5$, $0.8$ and $1$, they are clearly visible in the case of $\gamma^\text{init}=0.2$. This is confirmed by observing the behavior of the mean density sum (red line in Fig. \ref{fig:initexp}b), whose final value settles around \SI{10}{\percent} lower than the other three cases.

In addition to the intrinsic uncertainty on the density profile that derives from the presence of loosely constrained modes, the lower value for $\gamma^\text{init}=0.2$ could be partially explained as being due to the algorithm struggling during the \emph{shape identification} stage (see Sec. \ref{subsec:monitor}). This behavior can be discerned by monitoring the value of the oversampling parameter along the reconstruction process, reported as a yellow curve in Fig. \ref{fig:initexp}b. For the higher values of $\gamma^\text{init}$ the correct shape is quickly identified and, as a consequence, the oversampling value reaches its final value of $35$ within the first $20$ generations. In the lowest gamma value case (solid line) the convergence towards the correct shape requires around $60$ generations. The long-lasting larger support for $\gamma^\text{init}=0.2$ implies the presence of additional and stronger minima of the error function that the algorithm had to overcome during the retrieval (see Fig. \ref{fig:supports}). In addition, the algorithm had less ``time'' to optimize the reconstruction once that the correct shape is finally identified. This slow convergence is mostly caused by a threshold value $\tau=0.03$ of the SW algorithm (see Table \ref{tab:params}) set too low for the case of $\gamma^\text{init}=0.2$. However, the capability to converge in all the reported cases, with strongly different density normalizations, by using the same set of algorithm's parameters (and especially those concerning the support update) is a further proof of the resilience of SPRING.

A further demanding task for imaging algorithms is dealing with low-brightness diffraction data. In such cases, iterative algorithms tend to become unstable, especially when dealing with large samples, and the extraction of high-resolution features of the sample becomes particularly challenging. 
In Sec. 3.1 of the Supplemental Material SPRING is tested and compared with conventional methods on this kind of diffraction data.

\subsubsection{Complex-valued density} \label{subsec:complex}

\begin{figure*}
  \centering
  \includegraphics[width=0.85\linewidth]{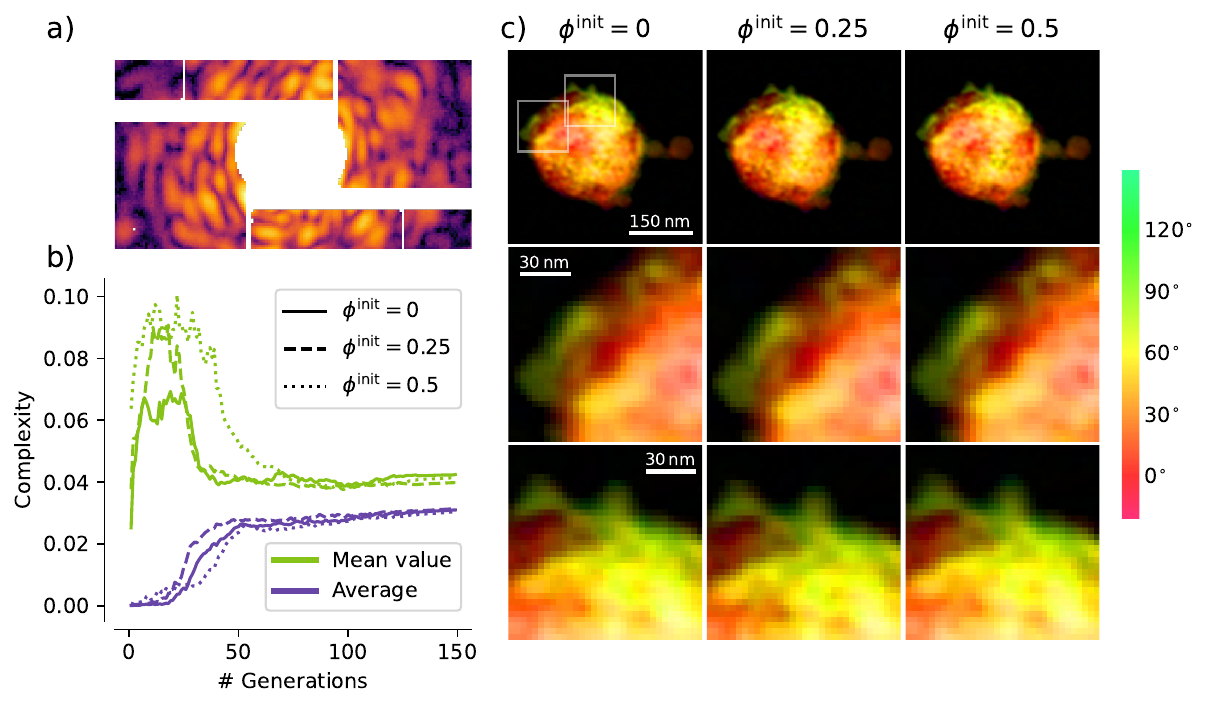}
  \caption{Effects of different phase values assigned to the starting guesses on the reconstruction outcomes for different values of $\phi^\text{init}$. In a) a cutout of the experimental data, extracted from Dataset A, shows asymmetries in the intensity distribution, which is a sign of phase contrast induced by the sample density. In b) the complexity indicator $\Gamma$ is plotted as function of the number of generations, both reporting its mean value within the population of reconstructions (green) and its value calculated on the average reconstruction (purple). In c) the final reconstruction outcome is reported for the three cases, with the phase information encoded in the color map shown on the side.}
  \label{fig:initphi}
\end{figure*}

The refractive index of materials in the X-ray regime, and in particular its real part, can be fairly approximated to $1$  \cite{paganin2006coherent}. The deviations from $1$ are often expressed in terms of $\delta = n - 1$ where $\delta \lesssim 10^{-2}$. 

However, the situation changes when the sample is composed of different materials with substantially different $\delta$ values at the probing photon energy. In this case, different phase shifts are induced in the light depending on the position in the sample, creating a phase contrast in the scattered field \cite{colombo2023imaging,colombo2022scatman}. This phase contrast affects not only the phases of the far field, but reshapes also the intensity distribution. CDI is, thus, inherently sensitive to phase contrast in the sample.
There are particular wavelengths where the small $\delta$ approximation does not hold, namely at photon energies in the vicinity of the core-excitation energies of atoms (a resonant condition) \cite{rupp2020imaging}.

In this regime, the $\delta$ coefficient for a specific element has a sharp fluctuation, thus increasing its phase contrast with other non-resonant materials in the sample. 
This is the case for mixed xenon-argon clusters imaged at 700eV, close to the 3d electronic resonance of xenon. An example of a diffraction pattern acquired in these conditions is reported in Fig. \ref{fig:initphi}a. In this close-up view of the central part of the pattern it is already possible to recognize non-centrosymmetric features, which are a sign of different phase shifts induced in the field by the sample. Those phase shifts are then reconstructed in real space as complex-valued densities.

When small phase shifts ($\ll \pi$) are expected in the reconstructed data, the positivity of the real part of the density is constrained along with the support function (see Appendix \ref{subsubsec:positivity}). This is achieved in SPRING by setting the $\chi$ parameter to $0.5$ (see Table \ref{tab:params}). Furthermore, it is convenient to initialize the densities as real-valued distributions. This is achieved by setting the SPRING parameter $\phi^\text{init}=0$ (see Appendix \ref{subsec:init}).

Instead, when relevant phase shifts are expected in the reconstruction, the phase retrieval problem becomes more challenging. First of all, the \emph{positivity} constraint of the real part of the reconstruction cannot be enforced \cite{marchesini2007invited}. This means that the $\chi$ SPRING's parameter has to be set to $0$. Furthermore, fully real-valued starting densities, initialized in SPRING by setting $\phi^\text{init}=0$, are sub-optimal starting conditions.

Fig. \ref{fig:initphi} reports the outcome of three MPR reconstructions with $\phi^\text{init}=0$ (completely positive real-valued initialization), $\phi^\text{init}=0.25$ (real-space starting phases range from $-\nicefrac{\pi}{4}$ to $\nicefrac{\pi}{4}$) and $\phi^\text{init}=0.5$ (real-space starting phases range from $-\nicefrac{\pi}{2}$ to $\nicefrac{\pi}{2}$). In all three reconstructions, \emph{unconstrained modes} are controlled by setting $C_a=0.3$ (see Appendix \ref{subsubsec:crossavg}).

The amount of phase shift in the reconstruction as function of the generations is tracked by the \emph{complexity} indicator $\Gamma$ (see Eq. \eqref{eq:complexity}), with values $0 \leq \Gamma \leq 1$. The value $\Gamma=0$ indicates a completely real-valued density. Its behavior as function of the generations is reported in Fig. \ref{fig:initphi}b. See Appendix \ref{subsec:missint} for further details and discussion.

Considering the mean value of the complexity in the population (green line), its initial value can be directly correlated to the range of the starting phases $\phi^\text{init}$, with higher values of the latter corresponding to greater values of the former. However, after an initial rise in the complexity value for all the three cases during the first generations of the algorithm, its value quickly drops and settles during the \emph{shape definition} stage at around $0.045$, consistently for the three different settings of $\phi^\text{init}$. The same consistency can be observed for the complexity value calculated on the average reconstruction (blue line in Fig. \ref{fig:initphi}b). Here, however, the complexity indicator stays close to $0$ as long as the \emph{shape definition} stage is reached, as widely differing densities in the population, despite being potentially strongly complex-valued, cancel out their phase deviations in the averaging operation. 

This consistency is directly reflected into the three reconstruction outcomes shown in Fig. \ref{fig:initphi}c. The overall density distribution and the respective phase values are fully matching. This is also true for the high resolution details highlighted by the zoomed-in cutouts of the reconstructions shown in the second and third row. It is worth mentioning that, for this diffraction pattern, the conventional IPR method couldn't reach proper convergence, as discussed in Fig. S5 of the Supplemental Material.

Thus, in most use-cases, starting densities can be initialized as completely real-valued ($\phi^\text{init}=0$), as different values of $\phi^\text{init}$ don't carry significant improvements on the final reconstruction outcome. This is a further demonstration of the stability of the MPR approach, which is capable of converging to the same solution with the same algorithm settings, but starting from very different, theoretically sub-optimal, conditions.

\section{Conclusions and Outlook}

We have presented an algorithmic framework to deal with the phase retrieval problem in single-particle single-shot Coherent Diffractive Imaging, called SPRING. SPRING implements the Memetic Phase Retrieval (MPR) \cite{colombo2017facing} method, which combines the widespread iterative phase retrieval algorithms with operations inherited from Genetic Algorithms tailored for XUV and X-ray Free-Electron Laser (XFEL) diffraction data.

The performance of SPRING has been investigated and compared with standard methods on data acquired in different experimental contexts with the two most common types of scattering detectors at XFELs. 
For diffraction patterns acquired in optimal conditions, the quality of the imaging result is consistent for a wide range of algorithm settings, virtually removing the case-by-case tuning required by conventional approaches.
Furthermore, its limits are tested for situations that are typically challenging or untreatable for standard data analysis workflows, including large sample sizes, large regions of missing scattering intensities and starting conditions significantly far from the ideal ones.
Overall, we show how SPRING is remarkably robust against variations of the settings, consistent in terms of quality of the imaging result, and flexible with respect to the type and characteristics of the diffraction data.

Those features are a game-changer for single-particle experiments at XFEL facilities, where the acquisition conditions, like the brightness of the diffraction data and the sample size, are often sub-optimal, not completely under control and vary on a shot-to-shot basis. 
In fact, the low sensitivity of SPRING to the algorithm parameters makes it possible to reconstruct large fractions of a typical dataset with the same algorithm settings. This is a pivotal aspect for CDI at XFELs, where analyzable diffraction images are typically recorded during an experiment at a rate from one to a few patterns per second (depending on the hitrate, the repetition rate of the XFEL and the frame-rate of the detector), leaving hundreds of thousands of bright patterns to be analyzed at the end of the campaign.

While the core structure of SPRING is established, further developments and enhanced performances of the approach are expected in the future. Among the foreseen improvements, the main ones involve the implementation of additional and more performant iterative phase retrieval algorithms and more sophisticated strategies to deal with noisy and missing diffraction data. Furthermore, the MPR algorithm is fully compatible with an extension to three-dimensional diffraction datasets \cite{colombo2018high}: its implementation in SPRING and the corresponding performance will be investigated in detail in future work. 

A Python module, called \emph{spring}, is released as open-source software in the context of this publication. The \emph{spring} package implements a highly optimized version of the algorithm, as well as additional helper functions to aid scientists in dealing with the preparation of experimental data and the imaging results. 

Currently, CDI experiments at XFELs are successfully accomplished mainly with the involvement of highly experienced research groups, due to the technical challenges to address in the experiment as well as due to the intricate and complex procedures required for data analysis and imaging.
Thanks to its enhanced reliability and the ease of use of its software implementation, SPRING will play a main role in easing the data analysis process and making imaging experiments at XFELs more accessible to a wider, less specialized and more applied scientific community.

\begin{acknowledgments}
The experimental campaigns were conducted at SwissFEL (proposal 20211710) and European XFEL (proposal 4396). 
We acknowledge the Paul Scherrer Institute, Villigen, Switzerland for provision of free-electron laser beam time at the Maloja instrument of the SwissFEL ATHOS branch and thank the SwissFEL staff for their assistance.
We acknowledge the European XFEL in Schenefeld, Germany, for the provision of x-ray free-electron laser beam time at the SQS instrument and thank the EuXFEL staff for their assistance. 
We acknowledge Olle Björneholm (Uppsala University), Julia Kojoj (Stockholm University) and Paul Zieger (Stockholm University) for their help and support during the European XFEL experimental campaign.
We acknowledge Dr. Adam Summers and Prof. Artem Rudenko for providing the aerodynamic lens setup at our disposal at SQS. 
AC and DRu acknowledge the Swiss National Science Foundation (via grant no. 200021E\_193642, grant no. 200021-232306, and the NCCR MUST) and ETH Zurich (via collaborative grant 23-2ETH-050) for financial support.
MP, OV, and MB further acknowledge the Research Council of Finland for financial support (including projects 326291, 330118, and 341288). 
TF acknowledges funding by the Deutsche Forschungsgemeinschaft within CRC 1477 "Light-Matter Interactions at Interfaces" (project number 441234705). PHWS acknowledges support from the Swedish Research Council through project 2018-00740.
FRNCM acknowledges the Swedish Research Council (2018-00234 and 2019-06092) and the Carl Tryggers Stiftelse för Vetenskaplig Forskning (CTS 19-227). JAS acknowledges the Swedish Research Council (2023-06350), the Göran Gustafsson Foundation (2044) and the Carl Tryggers Stiftelse för Vetenskaplig Forskning (CTS 21-1427).
We thank the IT Services Group (ISG) of the Department of Physics at ETH Zurich for the excellent support and management of the computing hardware on which the \emph{spring} software has been developed and tested.\\
\end{acknowledgments}

\subsubsection*{Code and data availability} 
The \emph{spring} module is released as open source software at \url{https://gitlab.ethz.ch/nux/spring}. The documentation is available at \url{https://share.phys.ethz.ch/~nux/software/spring/main/docs/}. A version of the code and its documentation is archived at \url{http://doi.org/10.5905/ethz-1007-832}. The raw data recorded for the experiment at the European XFEL are available at \url{https://doi.org/10.22003/XFEL.EU-DATA-004396-00}.\\

\subsubsection*{Author contributions} 
AC, MS, AA, TF, LH, GK, KK, BL, ICP, JCSZ, KS, MLS, ZSu, PT, CFU, SW, AVW, MZ, CB and DRu actively contributed in the experimental campaign at the Maloja endstation of SwissFEL (proposal 20211710). The experiment was led by MS and DRu. AC, MS, KA, MB, RB, RD, SD, FRNCM, AM, PM, TM, YO, DRa, AS, JAS, BS, SU, ZSh, PHWS, OV, NW, TY, MM, DRu and MP actively contributed in the experimental campaign at the SQS instrument of the European XFEL (proposal 4396). The experiment was led by MP. 
AC and DEG designed the original scheme of the MPR method. AC led the following development of SPRING with the input and support of DRu. AC designed, implemented and optimized the \emph{spring} software and wrote its documentation. AC, MS and DRu worked out the performance assessment of SPRING. AC and DRu wrote the main body of the manuscript, with contribution and assistance from all coauthors.\\

\subsubsection*{Competing interests} 
The authors declare no competing interests.


\appendix

\section{The phase problem as an optimization problem}\label{subsec:optprob}

Starting from any spatial density $\rho(\vec{x})$, a new density $\rho_S(\vec{x})$ that is fully contained in the support function $S(x,y)$ can be created by the following operator $P_S$:
\begin{equation}\label{eq:ps}
\rho_S = P_S \rho = S \cdot \rho
\end{equation}
where $\cdot$ indicates the element-wise matrix multiplication, and the dependence on the coordinates $x$ and $y$ has been omitted for the sake of brevity.

Similarly, it is possible to produce a density $\rho_M$, whose Fourier modulus is equal to the square root of the experimental data, by applying the following operator $P_M$ to any given density $\rho$:
\begin{equation}\label{eq:pm}
\rho_M = P_M \rho  = \mathcal{F}^{-1} \left[  M \cdot \frac{\tilde{\rho}}{\left| \tilde{\rho}  \right|}  \right]
\end{equation}
where $\mathcal{F}^{-1}$ is the inverse FT operation, $M$ is the square root of the experimental intensities $I$, and $\tilde{\rho} = \mathcal{F}[\rho]$.
The two operators in Eq.\eqref{eq:ps} and Eq.\eqref{eq:pm} can be easily demonstrated to be projectors, as ${P_M}^2 = P_M$ and ${P_S}^2 = P_S$, and they are usually addressed as \emph{modulus projector} and \emph{support projector} \cite{kirian2020imaging, colombo2023imaging, marchesini2007invited}.

The solution to the phase problem can be written in terms of $P_M$ and $P_S$ as the density function $\rho_\text{sol}$ that satisfies both constraints (i.e. $P_M \rho_\text{sol} = \rho_\text{sol}$ and $P_S \rho_\text{sol} = \rho_\text{sol}$).
It is convenient to define the error function:
\begin{equation}\label{eq:error}
E_{S,M}[\rho] = \lVert M - \mathcal{F}\left[ P_S \rho \right] \rVert
\end{equation}
where $\lVert \cdot \rVert$ denotes the L$^2$ norm.  Eq. \eqref{eq:error} indicates how \emph{far} a density $\rho$, projected on the support function, is from the experimental Fourier amplitudes $M$. In theory, the solution to the phase problem $\rho_\text{sol}$ has error $E_{S,M}[\rho_\text{sol}]=0$. In practice, due to intrinsic noise in the experimental data, the error $E$ in Eq. \eqref{eq:error} cannot reach $0$, and the solution $\rho_\text{sol}$ has to be defined as the one that minimizes the error, i.e.:
\begin{equation}\label{eq:min}
\rho_\text{sol} := \left\{ \rho : E_{S,M}[\rho] = \underset{\rho}{\min}\,E_{S,M} \right\} 
\end{equation}

Eq. \eqref{eq:min} expresses the phase retrieval problem as an \emph{optimization problem}. Phase retrieval algorithms are, thus, \emph{optimization algorithms}, which aim at minimizing the \emph{optimization target}, i.e. the error value, defined in Eq. \eqref{eq:error}.

The conceptually simplest phase retrieval algorithm, Error Reduction (ER) \cite{fienup1982phase}, acts by applying in an alternate manner the two projection operators in Eq. \eqref{eq:ps} and Eq. \eqref{eq:pm}. A single \emph{iteration} of the ER algorithm that constrains a given support function $S$ and experimental moduli $M$ can be thus defined as $\text{ER} = P_M P_S$. Its action on a density $\rho$ gives as a result a new density $\rho' = P_M P_S \rho$. 

The ER algorithm can be demonstrated to be a \emph{steepest-descent} method for the error $E$ because the error value is always reduced with iterations, i.e. $E[\text{ER}^{\iter+1} \rho] \leq E[\text{ER}^{\iter} \rho]$. This renders the ER algorithm particularly \emph{stable}, because it is guaranteed to converge to an optimum of the error value. It has however the drawback of getting stuck in local minima.

\section{Memetic operators}\label{sec:memeticops}

This section describes in a detailed way the SPRING implementation of the different operations performed by the MPR approach (see the overview in Fig. \ref{fig:flowchart}). The core of MPR \cite{colombo2017facing} is built around three operators inherited from \emph{Memetic algorithms} \cite{moscato2002memetic}. The \emph{crossover}, responsible for the creation of a new population of candidate solutions from the existing one, is discussed in Appendix \ref{subsec:crossover}. The \emph{self-improvement}, where conventional iterative phase retrieval algorithms are executed for each reconstruction in the population, is presented in Appendix \ref{subsec:improve}. The \emph{Selection}, where only the reconstructions with the lowest error value are selected to survive to the next algorithm generation, is described in Appendix \ref{subsec:selection}. Details on the initialization of the starting reconstructions is given in Appendix \ref{subsec:init}. An overview of the parameters that can be tuned in MPR is given in Appendix \ref{subsec:algparams}, followed by further considerations and implementation details on less central but still relevant operations implemented in the method.

In this section the following formalism is used. A reconstruction is indicated as $\rec$, which comprises the complex-valued matrix $\rec \mem \den_{ij}$ containing the density values and the boolean-valued matrix $\rec \mem \sup_{ij}$ that defines the corresponding support function. The Fourier Transformation (FT) of the density matrix is indicated as $\rec \mem \tilde{\den}_{ij}$. The density $\den$ and the support $\sup$ have dimension $N_p \times N_p$, equal to the size of the matrix containing the experimentally measured intensities $I_{ij}$ and the corresponding experimental modulus $M_{ij} = \sqrt{I_{ij}}$.

A population of $P$ reconstructions is reported as $\left\{ \rec^{p} \right\}$, with $p=1,...,P$. The density values of the reconstruction with index $p$ in the population is thus indicated as $\rec^{p} \mem \den_{ij}$.

\subsection{Crossover} \label{subsec:crossover}

\begin{figure*}
\begin{minipage}{\linewidth}
\begin{algorithm}[H]
\caption{The crossover routine}
\label{alg:crossover}

\begin{algorithmic}[1]
\Function{Crossover}{$\left\{ \rec^{p} \right\}$}

	\State $\left\{ \rec_\text{new}^{p} \right\} \gets \Call{Copy}{\left\{ \rec^{p} \right\}}$ \label{l:crossover:init} \algcom{Initialize the new population of reconstructions $\left\{ \rec_\text{new}^{p} \right\}$ as a copy of the old one.}

	\For{$p=1,...,P$}	\label{l:crossover:for} \algcom{Loop on the single individuals of the new population}
	
		\State $\rec^a, \rec^b, \rec^c  \gets \Call{RandomChoice}{\left\{ \rec^{p} \right\}}$ \label{l:crossover:extract} \algcom{Extract three random individuals from the parent population}
		
			\State $\rec^a, \rec^b, \rec^c  \gets \Call{Reshift}{ \left\{\rec^a, \rec^b, \rec^c \right\}, \rec^{p}}$ \label{l:crossover:reshift} \algcom{Transform the three extracted \emph{parents} such that they overlap in real space with the reconstruction $\rec^{p}$ }
		\State $C^\text{map} \gets \Call{CrossoverMap}{t_\text{min}, t_\text{max}, C_p}$\label{l:crossover:crossmap} \algcom{Create the binary crossover map $C^\text{map}$}
				
		\ForAll{$i,j$} \label{l:crossover:mixloop} \algcom{Loop over the matrix coordinates $i,j$}
			\If{$C_{ij}^\text{map} = 1$}

          \State $\rec_\text{new}^{p} \mem \tilde{\rho}_{ij} \gets \rec^a \mem \tilde{\rho}_{ij} + C_w \times \left( \rec^b \mem \tilde{\rho}_{ij} - \rec^c \mem \tilde{\rho}_{ij} \right)$ \label{l:crossover:mix}	 
          
          \algcom{Combine the densities of reconstructions $\rec^a$, $\rec^b$ and $\rec^c$ where $C_{ij}^\text{map} = 1$}
			\EndIf
			
			\State	$\rec_\text{new}^{p} \mem \sup_{ij} \gets \rec^a \mem \sup_{ij}  \textbf{ OR } \left( \rec^b \mem \sup_{ij} \textbf{ AND } \rec^c \mem \sup_{ij}   \textbf{ AND } \rec^{p} \mem \sup_{ij} \right)$ \label{l:crossover:mixsupp}
			
			 \algcom{Combine the supports of reconstructions $\rec^a$, $\rec^b$, $\rec^c$ and $\rec^p$ with logical operators}			
			
		\EndFor

	\EndFor
		
\State \Return $\left\{ \rec_\text{new}^{p} \right\}$ \algcom{The newly created population is returned as output of the \textsc{Crossover} function}
\EndFunction
\end{algorithmic}
\end{algorithm}
\end{minipage}
\end{figure*}

The crossover operation is a central step of an evolutionary algorithm. It is responsible for creating a new population from the pre-existing one, enabling the exploration of the parameters space. In the crossover step, reported in Algorithm \ref{alg:crossover}, a new population of reconstructions $\left\{ \rec^{p}_\text{new} \right\}$ is formed by combining individuals in the original population $\left\{ \rec^{p} \right\}$.

The new density values $\rec^{p}_\text{new} \mem \den_{ij}$ are filled by following the crossover scheme employed in Differential Evolutionary algorithms \cite{storn1997differential}. Some values of the new density matrix $\rec^{p}_\text{new}\mem\den_{ij}$  are directly copied from the corresponding element in the \emph{parent} population $\rec^{p}\mem\den_{ij}$.
The others are instead created by combining three randomly-chosen individuals in the original population $\left\{ \rec^{p} \right\}$. This operation is performed by the routine \textsc{RandomChoice} at line \ref{l:crossover:extract}, which returns three randomly-selected reconstructions from the \emph{parent} population, $\rec^a, \rec^b, \rec^c$ . The three extracted \emph{individuals} are different from each other and from $\rec^{p}$.
The values of the matrix of $\rec_\text{new}^{p}\mem\den_{ij}$ are set as a combination of the three ``parent'' densities $\rec^a\mem\den_{ij}, \rec^b\mem\den_{ij}, \rec^c\mem\den_{ij}$ following the formula at line \ref{l:crossover:mix} of Algorithm \ref{alg:crossover}.

There are two important preparatory steps for the actual mixing of the density values.
The first one is accomplished by the function \textsc{Reshift} at line \ref{l:crossover:reshift}, which deals with the ambiguities intrinsic of the phase retrieval problem. In fact, there exists a set of transformations of a density $\rho$ that only affect the phases of its Fourier representation, $\arg(\tilde{\rho})$. These transformations are:
\begin{enumerate}
\item Spatial translation: $\rho(\vec{x}) \rightarrow \rho(\vec{x} + \Delta \vec{x})$
\item Complex conjugation and coordinates inversion: $\rho(\vec{x}) \rightarrow \rho^*(-\vec{x})$
\item Addition of a global phase factor: $\rho(\vec{x}) \rightarrow e^{i \phi} \rho(\vec{x})$
\end{enumerate}
and any combination of those. 
These transformations have no effect on the modulus in Fourier representation, which is the only accessible experimental information on the sample. The direct implication is that a density $\rho$ and all its transformations listed above are equivalent candidate solutions to the phase retrieval problem.
These ambiguities must be resolved before any mathematical operation that involves multiple densities, like the crossover mixing (line \ref{l:crossover:mix} in Algorithm \ref{alg:crossover}) and the calculation of the average density. 

The operation is performed by the routine \textsc{Reshift} at line \ref{l:crossover:reshift} of Algorithm \ref{alg:crossover}. It requires a set of reconstructions to be transformed, in this case $\left\{ \rec^a, \rec^b, \rec^c \right\}$, and a reference reconstruction, in this case  $\rec^{p}$. The function then applies a combination of the transformations listed above to the input densities and supports such that they are as similar as possible to the reference density. This operation can be done efficiently by making use of cross-correlation functions, as detailed in section 2.2.7 of Ref. \cite{colombo2023imaging}, and as further discussed in Appendix \ref{subsec:reshift}.

\begin{figure}
  \centering
  \includegraphics[width=0.8\linewidth]{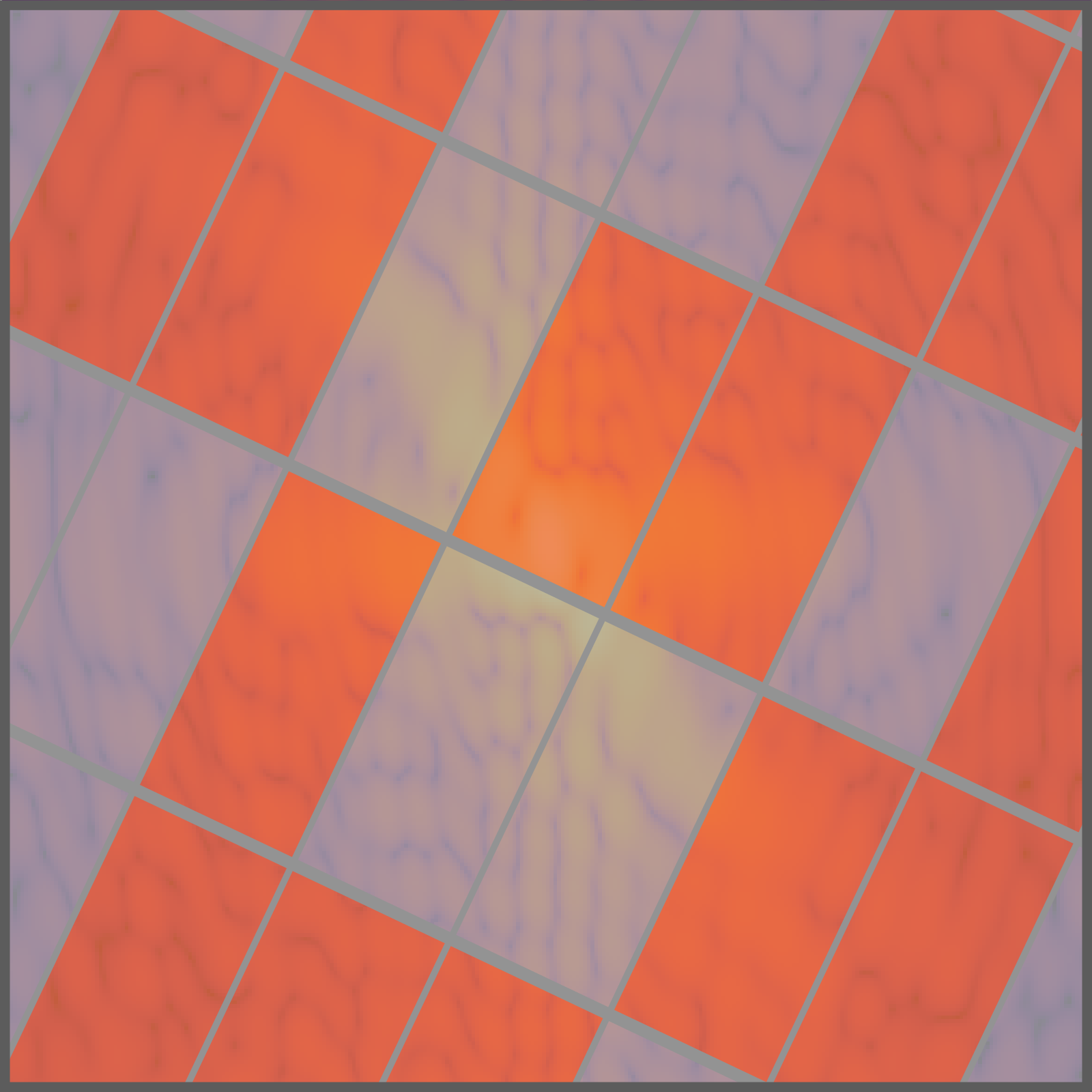}
  \caption{Example of a crossover map. The domain is divided into tiles of random height and width. A random shift in space, as well as a random rotation, are then applied to the tiles. Tiles are then elected for the crossover operation with a probability $C_p$. Coordinates of the matrix that belongs to the elected tiles are then subjected to the crossover mixing.}
  \label{fig:crossmap}
\end{figure}

A second preparatory step, accomplished by the \textsc{CrossoverMap} function at line \ref{l:crossover:crossmap} of Algorithm \ref{alg:crossover}, is to define which elements in the matrix $\rec_\text{new}^{p}\mem\tilde{\den}_{ij}$, i.e. the Fourier representation of the density matrix $\rec_\text{new}^{p}\mem\den_{ij}$, are modified by the mixing operation. As exemplified in Fig. \ref{fig:crossmap}, the Fourier domain is divided into rectangular tiles of size $t_x$ and $t_y$. The values of $t_x$ and $t_y$ are randomly chosen in the range $\left[ t_\text{min}, t_\text{max} \right]$, and given as parameters to the function \textsc{CrossoverMap}. The value of $t_\text{min}$ is typically on the order of a few pixels (e.g. $t_\text{min}=4$), while $t_\text{max}$ can assume values corresponding to a relatively large fraction of the linear dimension $N_p$ of the total matrix, for example $t_\text{max} = \nicefrac{N_p\,}{8}$. The tiles domain is then randomly shifted and rotated, as visible in Fig. \ref{fig:crossmap}. Afterwards, tiles are elected as ``crossover regions'' with a probability $C_p$, given as a parameter. 
The output of the \textsc{CrossoverMap} function is then a binary matrix $C^\text{map}$ of size $N_p \times N_p$, whose values are set to $1$ only at those coordinates $i,j$ belonging to tiles that have been selected for the crossover operation. The tiling of Fourier domain enables mixing of the information during the crossover operation without destroying the local correlations existing in the FT of the density. These correlations derive from the limited spatial extension of the sample and are a consequence of the \emph{oversampling} method. For further information, please refer to Ref. \cite{colombo2023imaging} and Ref. \cite{miao2003phase}.

It is finally possible to calculate the new values for the new density $\rec_\text{new}^{p}\mem\tilde{\den}_{ij}$. The operation is performed at line \ref{l:crossover:mix} only for those coordinates $i,j$ where the crossover map $C_{ij}^\text{map} = 1$. The mathematical operation depends on a parameter $C_{w}$, typical of Differential Evolution, called ``crossover weight'' \cite{storn1997differential}. 

Additionally, a new population of support functions has to be created. The mixing operation is performed here for all the entries in the new support matrix $\rec_\text{new}^{p}\mem\sup_{ij}$, without any tiling operation, by mixing the support information of the three selected densities, i.e.  $\rec^a\mem\sup_{ij}$, $\rec^b\mem\sup_{ij}$ and $\rec^c\mem\sup_{ij}$. 
As long as the support matrix is a binary function, the mixing operation at line \ref{l:crossover:mixsupp} of Algorithm \ref{alg:crossover} is performed via binary operator. 
In practical terms, the new support function is inherited from $\rec^a$, with the addition of those regions included in the support function of all the other parents $\rec^b$,$\rec^c$ and $\rec^p$.

Once the new densities and supports are created for all reconstructions in the new population $\left\{ \rec_\text{new}^{p} \right\}$, the real-space representation of the new densities $\rec_\text{new}^{p}\mem\den$ is calculated via an inverse FT of $\rec_\text{new}^{p}\mem\tilde{\den}$ and the new population is returned as output.

\subsection{The self-improvement} \label{subsec:improve}

\begin{figure*}
\begin{minipage}{\linewidth}
\begin{algorithm}[H]
\caption{The self-improvement routine}
\label{alg:improve}

\begin{algorithmic}[1]
\Function{Improve}{$\left\{ \rec^{p} \right\}$,$\left\{ \rec_\text{new}^{p} \right\}$ }

		\State $\alpha^\text{best} \gets \sum_{i,j} \rec^\text{best} \mem \sup_{ij}$ \label{l:improve:areabest} \algcom[0.6]{Calculate the area $\alpha^\text{best}$ of the reconstruction with the lowest error in $\left\{ \rec^{p} \right\}$.}
		
		\State $\alpha^\text{all} \gets \sum_{i,j} \, \bigcup\limits_{p=1}^{P} \rec^{p} \mem \sup_{ij}$ \label{l:improve:areaall} \algcom[0.6]{Calculate the area $\alpha^\text{all}$ of the union of all the supports in $\left\{ \rec^{p} \right\}$}
		
		\State $\alpha^\text{eval} \gets \frac{g}{G} \alpha^\text{best} + (1-\frac{g}{G}) \alpha^\text{all}$ \label{l:improve:areaeval} \algcom[0.6]{Calculate the area $\alpha^\text{eval}$ for the error evaluation of the reconstruction as a linear combination of $\alpha^\text{best}$ and $\alpha^\text{all}$, with a coefficient dependent of the current generation $g$}

		\ForAll {$\rec \in  \left\{ \rec^{p} \right\} \cup \left\{ \rec_\text{new}^{p} \right\}$} \label{l:improve:popfor} \algcom[0.5]{For all the reconstructions in the two populations}
		
			\Repeatn{$R-2$}	\label{l:improve:repfor}
				\State $\rec\mem \den \gets \Call{Sequence}{\rec\mem\den, \rec\mem\sup, M}$  \label{l:improve:sequence} \algcom[0.5]{Improve the density with a sequence of IAs}	
				\State $\rec\mem \sup \gets \Call{UpdateSupport}{\rec\mem\den, \sigma, \tau}$	\label{l:improve:sw}	\algcom[0.5]{Shrink the support function via the Shrink-wrap algorithm}			
			\EndRepeatn					
			
			\State $\rec\mem \den \gets \Call{Sequence}{\rec\mem\den, \rec\mem\sup, M}$ 
			\State $\rec\mem \sup \gets \Call{UpdateSupportSize}{\rec\mem\den, \sigma, \alpha^\text{best}}$\label{l:improve:swbest}	 
			
     		 \algcom[0.5]{Update the support to match a size equal to $\alpha^\text{best}$.}
     		 
			\State $\rec\mem \den \gets \Call{Sequence}{\rec\mem\den, \rec\mem\sup, M}$ 
					\State $\rec\mem \sup \gets \Call{UpdateSupportSize}{\rec\mem\den, \sigma, \alpha^\text{eval}}$\label{l:improve:sweval}

			\algcom[0.5]{Update the support to match a size equal to $\alpha^\text{eval}$.}

			\State $\rec\mem \den \gets \Call{EvalSequence}{\rec\mem\den, \rec\mem\sup, M}$ \label{l:improve:sequenceeval} \algcom[0.5]{Improve the density using only the Error Reduction algorithm}
			
			\State $\rec\mem \mathcal{E} \gets \Call{GetError}{\rec\mem\den, \rec\mem\sup, M}$ \label{l:improve:eval} \algcom[0.5]{Calculate the reconstruction error}
			
			\State $\rec\mem \sup \gets \Call{UpdateSupport}{\rec\mem\den, \sigma, \tau}$		\algcom[0.5]{Shrink the support function via the Shrink-wrap algorithm}

		\EndFor

		\State \Return $\left\{ \rec^{p} \right\}$,$\left\{ \rec_\text{new}^{p} \right\}$ \algcom[0.6]{Return the populations of optimized reconstructions}

\EndFunction 
\end{algorithmic}
\end{algorithm}
\end{minipage}
\end{figure*}

This step, commonly defined as \emph{self-improvement} in the context of memetic algorithms \cite{moscato2003gentle}, is responsible for the local optimization of the candidate solutions. In the framework of phase retrieval, this local optimization is accomplished by iterative phase retrieval algorithms (IAs) \cite{marchesini2007invited} in order to improve the densities $\rec\mem\den$ and the Shrink-wrap algorithm (SW) \cite{marchesini2003x} as regards their respective support functions $\rec\mem\sup$. The structure of the \emph{self-improvement} step is reported in Algorithm \ref{alg:improve}.
The alternation of IAs and SW is repeated $R$ times.

For the first $R-2$ repetitions (see line \ref{l:improve:repfor}), the densities are improved via IAs through the function \textsc{Sequence}, and the support functions are updated via the SW method through the function \textsc{UpdateSupport}. For the last two repetitions, at line \ref{l:improve:swbest} and line \ref{l:improve:sweval}, the update of the support is instead performed by the function \textsc{UpdateSupportArea} in place of the standard \textsc{UpdateSupport}. The \textsc{UpdateSupportArea} function ensures that the new support function has a total area (i.e. number of coordinates where its value is $1$) corresponding to the input value $\alpha$.

The first call of \textsc{UpdateSupportArea} (line \ref{l:improve:swbest}) updates all the supports to a target area $\alpha^\text{best}$, i.e. the area of the support corresponding to the best reconstruction (i.e. the one with the lowest error value), calculated at line \ref{l:improve:areabest} of Alg. \ref{alg:improve}. The role of this function call is to push away from stagnation those reconstructions that are stuck in a local minimum where the support area is significantly larger than the target.

The second call of \textsc{UpdateSupportArea} (line \ref{l:improve:sweval}) sets the support area to a value $\alpha^\text{eval}$. The quantity $\alpha^\text{eval}$ is calculated as a linear combination of $\alpha^\text{best}$ and of the area of the union of all supports $\alpha^\text{all}$ (see line \ref{l:improve:areaall}). The coefficient of the linear combination $\nicefrac{g}{G}$ depends on the current algorithm generation $g=1,...,G$, tuning $\alpha^\text{eval}$  closer to $\alpha^\text{all}$ at the beginning of the reconstruction and closer to $\alpha^\text{best}$ towards the last generations (see line \ref{l:improve:areaeval}).
This second call ensures that all the reconstructions in the two populations have the same support area before the execution of the last algorithms sequence \textsc{EvalSequence} (line \ref{l:improve:sequenceeval}, where only the Error Reduction algorithm is executed) and the following error evaluation (line \ref{l:improve:eval}). The importance of this aspect and its central role in the MPR method is described in the following Appendix \ref{subsubsec:modshrink}.

Once that the error value is calculated via the \textsc{GetError} function, which calculates the quantity introduced in Eq. \eqref{eq:error_calc}, the support of the reconstructions are upgraded via the \textsc{UpdateSupport} call (standard Shrink-wrap algorithm) and the two \emph{improved} populations $\left\{ \rec^{p} \right\}$ and $\left\{ \rec_\text{new}^{p} \right\}$ are returned.

\subsubsection{The modified Shrink-wrap} \label{subsubsec:modshrink}

The functions \textsc{UpdateSupportSize} in Algorithm \ref{alg:improve} indicate a modified version of the Shrink-wrap algorithm for the update of the support function. 
This algorithm inherits the first step from the Shrink-wrap, i.e. it performs a gaussian smoothing, with standard deviation $\sigma$ given as parameter, on the input density $\rec\mem\den$. The standard Shrink-wrap then calculates the new support by thresholding  the smoothed density $\rho_\sigma$ at a value $\tau \times \max[\rho_\sigma]$. 

Conversely, this modified version ensures that the overall extension of the support, i.e. the number of coordinates where the density is allowed to be non-zero, equals to the quantity $\alpha$. The threshold $\tau$ is then automatically computed within the \textsc{UpdateSupportSize} based on the value $\alpha$ and the values of the smoothed density $\rho_\sigma$. 

This peculiar way of upgrading the support matrix ensures that each density $\rec\mem\den$ is then optimized by using a support constraint $\rec\mem\sup$ with equivalent ``strength''. This aspect is of fundamental importance for the following \textsc{Selection} step (see line \ref{l:mpr:select} of Algorithm \ref{alg:mpr}), where reconstructions in $\left\{ \rec^{p} \right\}$ and $\left\{ \rec_\text{new}^{p} \right\}$ are compared based on their error value, as discussed in the following section.

\subsection{The Selection step} \label{subsec:selection}

\begin{figure*}
\begin{minipage}{\linewidth}
\begin{algorithm}[H]
\caption{The selection step}
\label{alg:select}

\begin{algorithmic}[1]
\Function{Select}{$\left\{ \rec^{p} \right\}$,$\left\{ \rec_\text{new}^{p} \right\}$ }

		\For{$p=1,...,P$}	\label{l:select:repfor} \algcom[0.6]{Loop over all indexes $p$ in the reconstructions populations of size $P$}
			
			\If {$\rec_\text{new}^{p}\mem\mathcal{E}< \rec^{p}\mem\mathcal{E}$ } \label{l:select:geterror}	\algcom[0.6]{For each index $p$, the error of the reconstruction in the parent population $\rec^{p}$ and in the new population $\rec_\text{new}^{p}$ are compared}
				\State $\rec^{p} \gets \rec_\text{new}^{p}$	\label{l:select:replace} \algcom[0.6]{If the error of the reconstruction belonging to the new population is lower than the one of the parent population, replace the parent individual with the new one}
			\EndIf

		\EndFor
		\State \Return $\left\{ \rec^{p} \right\}$

\EndFunction 
\end{algorithmic}
\end{algorithm}
\end{minipage}
\end{figure*}

The implementation of the \textsc{Select} function at line \ref{l:mpr:select} of Algorithm \ref{alg:mpr} is reported in Algorithm \ref{alg:select}. It has the important role of comparing the performance of the reconstructions in $\left\{ \rec^{p} \right\}$ and $\left\{ \rec_\text{new}^{p} \right\}$ and upgrading the current population $\left\{ \rec^{p} \right\}$ accordingly. This operation is based on the evaluation of the error value, calculated at line \ref{l:improve:eval} of Alg. \ref{alg:improve}, for each reconstruction $\rec$ in the two populations. For each index $p$, the error values of the reconstruction in the original population $\rec^{p}$ and in the new one $\rec_\text{new}^{p}$ are compared (line \ref{l:select:geterror} of Alg. \ref{alg:select}).
The reconstruction $\rec^{p}$ is then replaced by $\rec_\text{new}^{p}$ if the error of the latter $\rec_\text{new}^{p}\mem \mathcal{E}$ has a lower value, otherwise $\rec_\text{new}^{p}$ is discarded. The upgraded population $\left\{ \rec^{p} \right\}$ is then returned by the function and a new generation of the MPR algorithm starts.

\begin{figure*}
  \centering
  \includegraphics[width=0.7\linewidth]{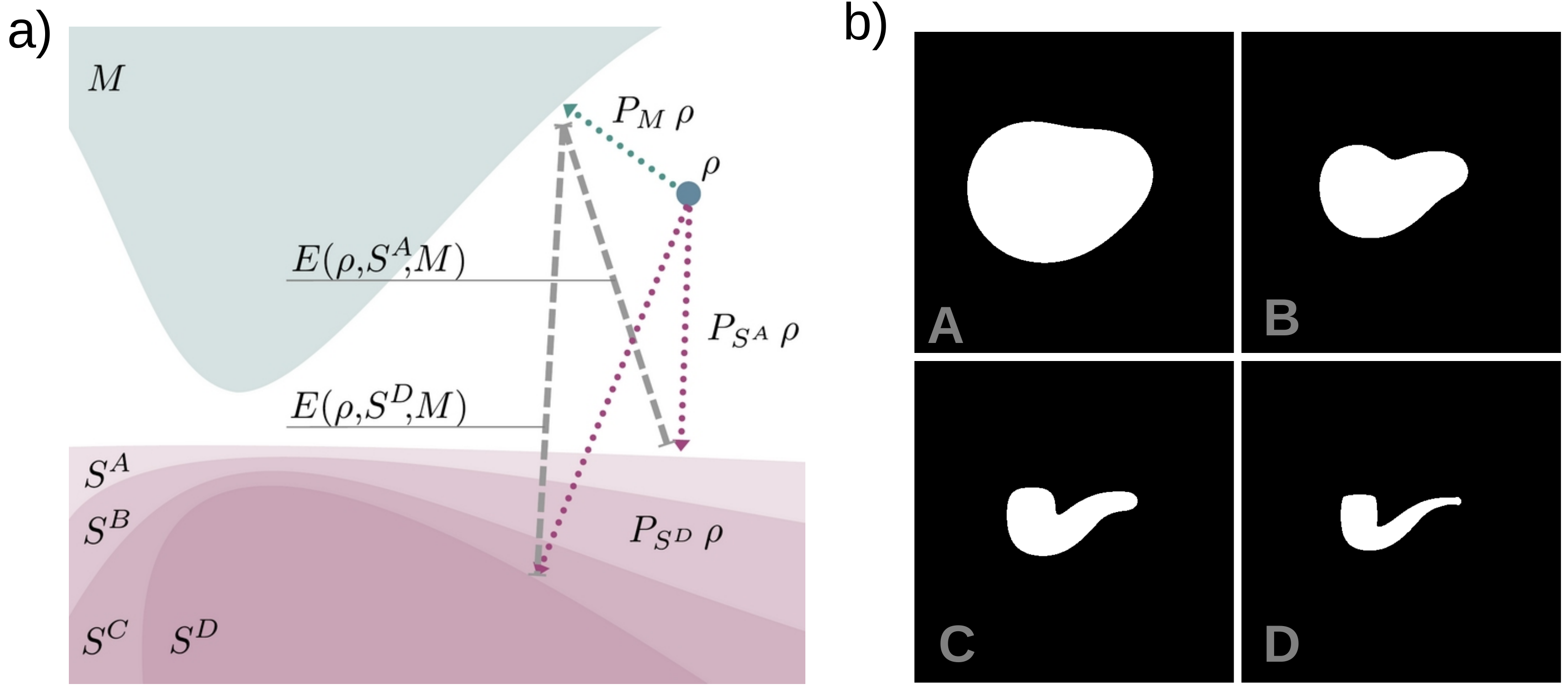}
  \caption{Effect of different support sizes on the evaluation of the error value. In a), graphical representation of the action of the two projectors, $P_M$ and $P_S$, on a generic density $\den$. The error value, indicated as a dashed gray line, is the distance between the two projections. For further discussion about the interpretation of the phase retrieval problem in terms of sets, please refer to Ref. \cite{marchesini2007invited}, \cite{kirian2020imaging} and \cite{colombo2023imaging}. In b) supports with decreasing sizes, from A to D, correspond to decreasing volumes of the sets  $S^A$ to $S^D$ in a). The dependency of the error value $E$ on different support sizes is also shown in a), with larger supports yielding smaller error values for the same density $\den$.}
  \label{fig:supports}
\end{figure*}

Despite its conceptual simplicity, the selection step has a peculiar aspect deriving from the nature of the phase retrieval problem, and has to be carefully handled as discussed in the following. 
The evaluation of the error involves the use of the two constraints required to ensure the existence of an unique solution to the phase problem, which are enforced by the two projector operators, $P_M$ and $P_S$, introduced by Eq. \eqref{eq:pm} and Eq. \eqref{eq:ps}, respectively. 

The action of the two projections on a generic density $\den$ is graphically represented in Fig. \ref{fig:supports}a, along with the error value $E$ which can be interpreted as the \emph{distance} between the two projections. Finding the solution to the phase problem can be also interpreted in an intuitive manner as the density $\den$ for which the distance between the projections on the two sets is the smallest.

The $P_M$ operator acts equivalently for all reconstructions, because the set $M$ is the same for all the densities (the experimental pattern $I_{ij}$ is known a-priori and does not change during the reconstruction procedure). 
The situation is substantially different for the case of the $P_S$ operator, which projects a given density $\den$ onto the set $S$. 
Each reconstruction in a population has its own support function which differs from the others. As a consequence, the projection operator $P_S$ acts differently for each reconstruction $\rec$.

This feature poses a problem when evaluating the error value of the reconstruction, which can be interpreted as a distance between the two sets as depicted in Fig. \ref{fig:supports}. In principle, each of the supports shown in Fig. \ref{fig:supports}b is enough to define a unique solution to the phase problem. However, a looser support function corresponds to a set $S$ in Fig. \ref{fig:supports}a with higher volume. Thus, reconstructions evaluated on a larger support function inherently have a smaller error value, as the distance between the sets $M$ and $S$ reduces. It is in general incorrect to compare the errors of two reconstructions with different support functions.

However, the comparability between the reconstructions can be restored by enforcing all the reconstructions $\left\{ \rec^{p} \right\}$ and $\left\{ \rec_\text{new}^{p} \right\}$ to have a set $S$ with equal volume, i.e. their support functions $\rec^{p}\mem\sup$ to be composed by the same number of pixels. This fundamental step is achieved by the \textsc{UpdateSupportSize} function described in section \ref{subsubsec:modshrink}, executed right before the last sequence of iterative algorithms \textsc{EvalSequence} in the \textsc{Improve} routine (at line \ref{l:improve:sequenceeval} of Alg. \ref{alg:improve}) and the error evaluation (line \ref{l:improve:eval} of Alg. \ref{alg:improve}).

\subsection{Initialization of the starting densities} \label{subsec:init}

The initialization of the starting guesses at the beginning of the reconstruction process, so far indicated by the \textsc{Initialize} function, plays a crucial role. On the one hand, it is desirable to create starting guesses as close as possible to the solution of the phase problem. On the other hand, the process has to be sufficiently randomized to enable a reasonably wide exploration of the parameters space. 
Despite that the initialization itself does not conceptually belong to the MPR method, the way in which densities are initialized may affect the final outcome of the reconstruction, and its description is useful to understand and evaluate the tests conducted on experimental data presented in Sec. \ref{sec:expdata}.

The initialization step employed here is based on filling a region of the real space with a number of spherical densities. The real space region is a square of size $\Sigma^\text{init}\times \Sigma^\text{init}$ pixels. This area is then filled by summing up a number $N^\text{init}$ of densities of projected spheres with different radii $R_n^\text{init}$. To each sphere, a global phase value $\phi_n^\text{init}$ can be assigned.

\begin{figure*}
  \centering
  \includegraphics[width=0.99\linewidth]{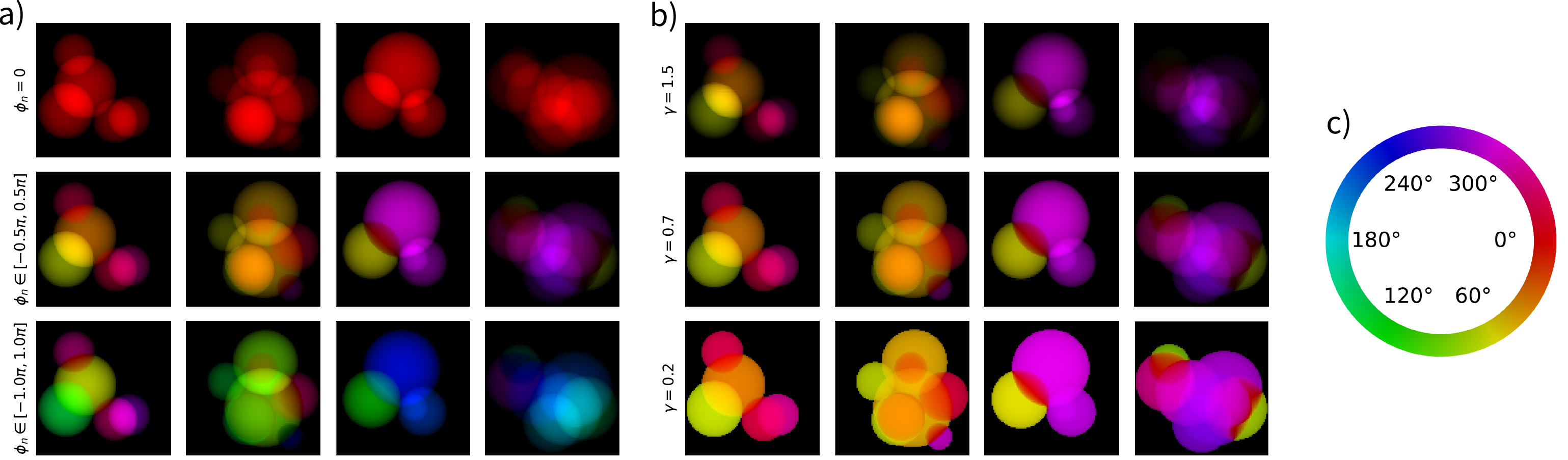}
  \caption{Examples of densities initialization. Densities are randomly initialized by filling a square area of the complex-valued matrix of spherical profiles with random sizes and positions. In a), for each row, four different densities are initialized with increasing values for the phases $\phi_n^\text{init}$. In b), different gamma values are applied to the densities. The phase of the complex-valued matrix are encoded following the color scheme shown in c).}
  \label{fig:init}
\end{figure*}

An example of this procedure is shown in Fig. \ref{fig:init}a, where four densities are stochastically initialized. In the first row, all the densities are real-valued, i.e. they all have a fixed global phase, in this case $\phi_n^\text{init}=0$ (see color map in Fig. \ref{fig:init}c). Here, densities are initialized in a square of edge $\Sigma^\text{init} = 100$ pixel. The number of spherical profiles is randomly chosen in the range $N^\text{init} \in \left[ 2, 8 \right]$. Their diameter is extracted in the range $D_n^\text{init} \in [20, 60]$ pixel.
The second and third rows of Fig. \ref{fig:init}a show, instead, the same initialization but with different ranges for the spheres phases, which are $\phi_n^\text{init} \in \left[-\nicefrac{\pi}{2}, \nicefrac{\pi}{2} \right]$ and $\phi_n^\text{init} \in \left[\pi, \pi \right]$ respectively.

The densities shown in Fig. \ref{fig:init}a correspond to the optical depth of spherical particles. These profiles can additionally be tuned by applying an exponentiation operation to the modulus of the overall density with exponent $\gamma^\text{init}$, i.e. $\left| \den \right|^{\gamma^\text{init}}$. All the densities in Fig. \ref{fig:init}a are produced with $\gamma^\text{init}=1$. Fig. \ref{fig:init}b shows instead the effect of different values for the initialization exponent $\gamma^\text{init}$, i.e. $1.5$, $0.7$ and $0.2$, applied to the densities of the second row in Fig. \ref{fig:init}a. The effect of $\gamma^\text{init}$ is equivalent to the \emph{gamma} value in image processing, where contrast can be increased ($\gamma^\text{init}>1$) or decreased ($\gamma^\text{init}<1$).

The last step performed in the initialization step concerns the normalization of the density. The Parseval's theorem states that the Fourier transform operation preserves the $L^2$ norm. Thus, the $L^2$ norm of the density $\den$ is fixed by the knowledge of the diffraction pattern $I$. In numerical terms, it means that $\sum_{ij} \left| \den_{ij} \right|^2 = \sum_{ij} \left| \text{FT}\left[ \den \right]_{ij} \right|^2 = \sum_{ij} I_{ij}$. So, if the full diffraction image $I_{ij}$ is known, the normalization is automatically constrained by the application of the modulus projector $P_M$ in Eq. \eqref{eq:pm}. 

However, one unavoidable limiting factor of real experimental data is that the inner part of the diffraction image is not recorded, due to the presence of a beam stopper or a hole in the detector to prevent from damages created by the highly intense transmitted light beam. This feature represents the main obstacle when attempting CDI on experimental data, as the phase retrieval algorithms have to also retrieve the missing intensities where the data is not available \cite{colombo2023imaging}. This aspect is discussed in detail in Sec. \ref{sec:expdata}. 

To aid phase retrieval algorithms in their optimization task, it is convenient attempt a normalization of the starting densities that is as close as possible to the correct one. Known values in the scattering image can be identified by a \emph{mask} matrix $\Xi$, with $\Xi_{ij}=1$ where the scattering values $I_ij$ are known, and $\Xi_{ij}=0$ otherwise. Each density $\den$ is then multiplied by a normalization factor:

\begin{equation}\label{eq:norm}
\nu  = \frac{\sum_{ij} \Xi_{ij} I_{ij}}{\sum_{ij} \Xi_{ij} \left| \text{FT}{\left[ \den \right] }_{ij}\right|^2}
\end{equation}
In practical terms, the full $L^2$ norm is inferred by comparing the \emph{partial} $L^2$ norms calculated only on the known pixels in the scattering data, where $\Xi_{ij}=1$.

\subsection{Summary of the parameters and further considerations}\label{subsec:algparams}

The parameters and terms introduced so far are summarized in Table \ref{tab:params}. This overview allows to further discuss some aspects that were only briefly mentioned in the previous section and require more discussion.

\begin{table*}[!]
\caption{Glossary of the most important parameters and keywords of the SPRING framework.}
\label{tab:params}
\small
\begin{tabularx}{\textwidth}{lcccX}
\toprule
Name & Type & Allowed values & Default & Description\\
\midrule
\midrule
\multicolumn{4}{l}{\textbf{Global} (see Sec. \ref{sec:MPR})} \vspace{0.3em}\\
$P$	& \textbf{int} & $\geq 4$& $128$ & 	Population size  \\
$G$	& \textbf{int} & $>1$ &	 $100$ & Number of generations  \\
$N_p$	& \textbf{int} & -- & -- & 	Linear dimension of the diffraction data  \\

\midrule
\midrule
\multicolumn{4}{l}{\textbf{Initialization} (see Appendix \ref{subsec:init})} \vspace{0.3em}\\
$\Sigma^\text{init}$ & \textbf{int}  & --& -- & 	Initial linear dimension of the support function \\
$N^\text{init}$ & \textbf{int} & $\geq 1$ & $5$ & 	Amount of spherical profiles that fills the initial support  \\
$D_n^\text{init} $ &  \makecell[t]{rand (\textbf{float})}& $[>0, \leq \Sigma^\text{init}]$ & $[0.2 \cdot \Sigma^\text{init}, 0.9 \cdot \Sigma^\text{init}]$ &  Range of diameters for the spherical profiles \\	
$\phi_n^\text{init}$ & \makecell[t]{rand (\textbf{float})} & $[\geq -\pi, \leq \pi]$  & $[0,0]$ &  Range of phase values for the spherical profiles \\	
$\gamma^\text{init}$ & \textbf{float} & $>0$ & $1$ &  Gamma value applied to the modulus of the initial density  \\

\midrule
\midrule
\multicolumn{4}{l}{\textbf{Crossover}} \vspace{0.3em}\\
$C_p$	& \textbf{float} & $ 0 < \cdot  \leq 1$ & $0.6$ & 	Crossover probability (see Appendix \ref{subsec:crossover})   \\
$C_w$  & \textbf{float} & $ 0 \leq \cdot  \leq 2$ & $0.4$ & 	Differential weight (see Appendix \ref{subsec:crossover})   \\
$C_a$  & \textbf{float} & $ 0 \leq \cdot  \leq 1$ & $0$ & 	Crossover averaging  (see Appendix \ref{subsubsec:crossavg})   \\

\midrule
\midrule
\multicolumn{4}{l}{\textbf{Support update}} \vspace{0.3em}\\
$\tau, \tau_\text{end}$ & \textbf{float} & $ 0 <  \cdot \leq 1$ & --, $\nicefrac{2}{3}\tau$ &  Threshold for the support update via the shrink-wrap algorithm   (see Sec. \ref{sec:phaseproblem})\\
$\sigma, \sigma_\text{end}$ & \textbf{float} & $ \geq 0$ & $2$, $0.5$ &  Smoothing amount for the support update via the shrink-wrap algorithm  (see Sec. \ref{sec:phaseproblem}) \\

\midrule
\midrule
\multicolumn{4}{l}{\textbf{Iterative algorithms}} \vspace{0.3em}\\
$R$	& \textbf{int} & $ \geq 2$  &	 $3$ & Number of repetitions of the main algorithms sequence (see Appendix \ref{subsec:improve})  \\
$\chi$	& \textbf{float} & $ 0 <  \cdot \leq 1$  & $0.5$ &  Fraction of the complex plane allowed for density values (see Appendix \ref{subsubsec:positivity}) \\
$\iter$	& \textbf{int} & $ >0 $  & $40$ &  Starting number of iterations of IA (HIO or RAAR) (see Appendix \ref{subsubsec:sequence}) \\
$\beta$ & \textbf{float} & $ 0 <  \cdot \leq 1$ & $0.9$ &  Beta parameter for HIO or RAAR algorithms (see Appendix \ref{subsubsec:sequence}) \\
$\iter_\text{ER}$	& \textbf{int} & $ >0 $  & $40$ &  Starting number of iterations of Error Reduction  (see Appendix \ref{subsubsec:sequence})\\
$\iter_\text{eval}$	& \textbf{int} & $ >0 $  & $40$ &  Number of iterations of Error Reduction for the evaluation sequence  (see Appendix \ref{subsubsec:sequence})\\

\end{tabularx}
\end{table*}%

\subsubsection{Changing values over generations} 
As the MPR generations progress it is useful to modify some parameters that affect the update of the support function and the execution of iterative algorithms. This is particularly true for the number of iterations of iterative algorithms (HIO and RAAR). In fact, a large number of iterations of these iterative algorithms (IAs) allows better exploration of the parameters space, suffering less stagnation issues compared to the ER algorithm. Their drawback is, however, that they often fail to keep the solution once they are close by \cite{marchesini2007invited}. A common practice is then to reduce their number of iterations over the reconstruction process. For these reasons, their number of iterations $\iter$ is set at the beginning, and it is linearly reduced to $\iter_\text{end}$ over the generations. By default, $\iter_\text{end}=0$ for HIO and RAAR.

A similar consideration applies to the update of the support function via the Shrink-wrap algorithm \cite{marchesini2003x}. There, both the threshold $\tau$ and the smoothing about $\sigma$ can be reduced along the reconstruction process to improve the quality of the reconstruction result. In this framework, the starting values of $\tau$ and $\sigma$ are then linearly changed as function of the generations to the final values $\tau_\text{end}$ and $\sigma_\text{end}$. By default, these values are set to $\tau_\text{end} = \nicefrac{2}{3} \tau$ and $\sigma_\text{end} = \SI{0.5}{\pixel}$.

\subsubsection{The algorithms sequence} \label{subsubsec:sequence} Iterative algorithms are executed as a \textsc{Sequence}, introduced in Eq. \eqref{eq:sequence}, which is repeated a number $R$ of times for each reconstruction $\rec$ at each generation $g=1,...,G$, and alternated with the routine \textsc{UpdateSupport} at each repetition (see Algorithm \ref{alg:improve}). While, in principle, several different combinations of algorithms can be used, the structure of the algorithms sequence is kept as simple as possible in this work.
As highlighted in Appendix \ref{subsec:improve}, two algorithms sequences are used:

\paragraph*{Main sequence} It is indicated as \textsc{Sequence}, and it is composed of some iterations of the HIO or RAAR algorithm (with the given value of the $\beta$ parameter), followed by some iterations of the Error Reduction algorithm. The starting number of IAs iterations is set by the parameter $\iter$ and the starting number of ER iterations is set by $\iter_\text{ER}$ (see Table \ref{tab:params}). Their number is linearly modified as the generations progress. Their final value is $0$ for IAs and $\iter + \iter_\text{ER}$ for the ER algorithm, i.e. only the Error Reduction algorithm is executed towards the end of the reconstruction procedure.

\paragraph*{Evaluation sequence}
The evaluation sequence is designed to steer the current reconstruction towards the closest local minimum of the error function, immediately preceding the error evaluation. For this reason, in the current implementation of MPR, it is composed only of a number of iterations $\iter_\text{eval}$ of the Error Reduction algorithm that is kept constant during the reconstruction process.

\subsubsection{Real-space constraints} \label{subsubsec:positivity} In addition to the support function, a further constraint can be added in real space. In literature, this constraint is indicated often as \emph{positivity} constraint. When applied, it restricts the real part of the density values $\den_{ij}$ to be non negative \cite{marchesini2007invited}. In general, the values $\den_{ij}$ are allowed to cover the complete complex set, i.e. their phases $\arg(\den_{ij}) \in \left[ -\pi, \pi \right]$, while $\arg(\den_{ij}) \in \left[ -\nicefrac{\pi}{2}, \nicefrac{\pi}{2} \right]$ if positivity is constrained. Within MPR, the positivity constraint is generalized by the parameter $\chi$, such that phases are, in general, constrained to the set $\left[ - \chi \,\pi, \chi \,\pi \right]$. The default value is $\chi = 0.5$, which is equivalent to the application of the typical positivity constraint. In the examples shown in this manuscript, $\chi$ is set to its default value if not explicitly indicated.

\subsubsection{Aligning the reconstructions} \label{subsec:reshift}

The \textsc{Reshift} operation, discussed in Appendix \ref{subsec:crossover}, is necessary to minimize the intrinsic ambiguities of the phase retrieval problem before performing operations that combine different densities, like the \emph{crossover} or the averaging. A generic density $\den$ can be transformed to match a \emph{reference density} $\den^\text{ref}$ by calculating the cross correlation function between the two, i.e. $\den \star \den^\text{ref}$. The coordinates of $\max \left[ \left|\den \star \den^\text{ref}\right|\right]$ (or, to be more precise, the coordinates relative to the center of the autocorrelation matrix) indicate the spatial shift that has to be applied to $\den$ to spatially overlap with $\den^\text{ref}$. The global phase that has to be added to $\den$ is, instead, the phase of this maximum value.
The ambiguity of complex conjugation and inversion, i.e. $\den(\vec{x}) \to \den^* (-\vec{x})$, is resolved by performing the same operations on the transformed version $\den^* (-\vec{x})$. Then, the one that gives the highest maximum value for the autocorrelation is kept.

This operation turns out to be safe and stable when the population of reconstructions is already converging towards a solution, while it may present some issues at the beginning of the reconstruction process. This is particularly true for large samples, where the relative amount of information lost in the central hole of the detector is relevant and renders the reconstruction process challenging. In fact, the dominant information that determines the outcome of the cross-correlation operation is the most intense one in Fourier space. Due to the isolated nature of the samples (necessary to satisfy the oversampling condition) the brightest region in the Fourier domain is the one at the center, corresponding to low momentum transfer, where experimental data is usually missing. The misestimate of these values by phase retrieval algorithms may result in misalignment of the reconstructions, and further amplifications of the corresponding artifacts (see also Appendix \ref{subsec:missint}). 

To mitigate this issue, the coordinates of the intensity matrix at which the experimental values are unknown are excluded from the calculation of the cross-correlation function. In particular, the cross-correlation between $\den$ and $\den^\text{ref}$ is calculated via the \emph{convolution theorem} as follows:
\begin{equation}\label{eq:crosscorr}
\den \star \den^\text{ref} = \mathcal{F}^{-1} \left[ \Xi \cdot \overline{\mathcal{F}[\den]} \cdot \mathcal{F}[\den^\text{ref}] \right]
\end{equation}
where $\Xi_{ij}$, called mask, has values $1$ where the intensities are known and $0$ otherwise.

\section{Dealing with missing intensities} \label{subsec:missint}

The unavoidable feature of CDI diffraction patterns at FELs is the lack of information in some regions of the reciprocal space. Those regions can be categorized into three types:

\begin{enumerate}
\item Regions in-between the detector modules. CDI detectors are constructed as an assembly of tiles, with some space between them to leave room for electronics. Those areas can be identified in the dark stripes of the patterns shown in Fig. \ref{fig:dataexamples}. 
\item Central region of the scattering image. A center hole allows the unscattered beam to pass through the detector without damaging it. This appears in the diffraction patterns as a large area of unknown intensities at the center of the detector with circular (Fig. \ref{fig:dataexamples}a) or rectangular (Fig. \ref{fig:dataexamples}d) shape.
\item Saturated pixels. Detector pixels are capable of recording radiation up to a maximum threshold value. Above this value, the pixel sensors no longer respond in a linear manner to the incoming radiation. Thus, those pixels have to be excluded from the retrieval procedure, as they do not carry correct information. Due to the decreasing scattering intensity with increasing momentum transfer, which is an intrinsic feature of diffraction images from isolated particles, saturated pixels are typically close to the detector center, further reducing the information recorded at low momentum transfer.
\end{enumerate}

While 1) and 2) are predetermined by design, 3) varies on a shot-to-shot basis. 
Some strategies are implemented in MPR to deal with these issues. 

With regards to point 1), iterative algorithms typically over-estimate the scattering signal in those regions, causing the appearance of artifacts in the reconstruction. A strategy to extract an upper bound for the missing intensities from the known ones has been developed. This is accompanied by the introduction of a modified version of the modulus projector $P_M$ that constrains this upper bound during the execution of iterative algorithms. This method greatly improves the quality of reconstructions belonging to Dataset B (Fig. \ref{fig:dataexamples}d). It is used in all of the tests that involve this dataset. It is worth underlining that this strategy is not strictly connected to the use of the MPR algorithm and can be directly implemented into standard imaging workflows. The implementation details and technical aspects of this strategy are reported in the following Appendix \ref{subsec:intbound}. 

The way in which MPR deals with the missing intensities in the central region of the detector corresponding to low momentum transfer, deriving from points 2) and 3), is instead strictly connected to the \emph{crossover} operation.
This topic is discussed in the following section.

\subsection{Constraining intensities at large scattering angles} \label{subsec:intbound}

\begin{figure*}
  \centering
  \includegraphics[width=\linewidth]{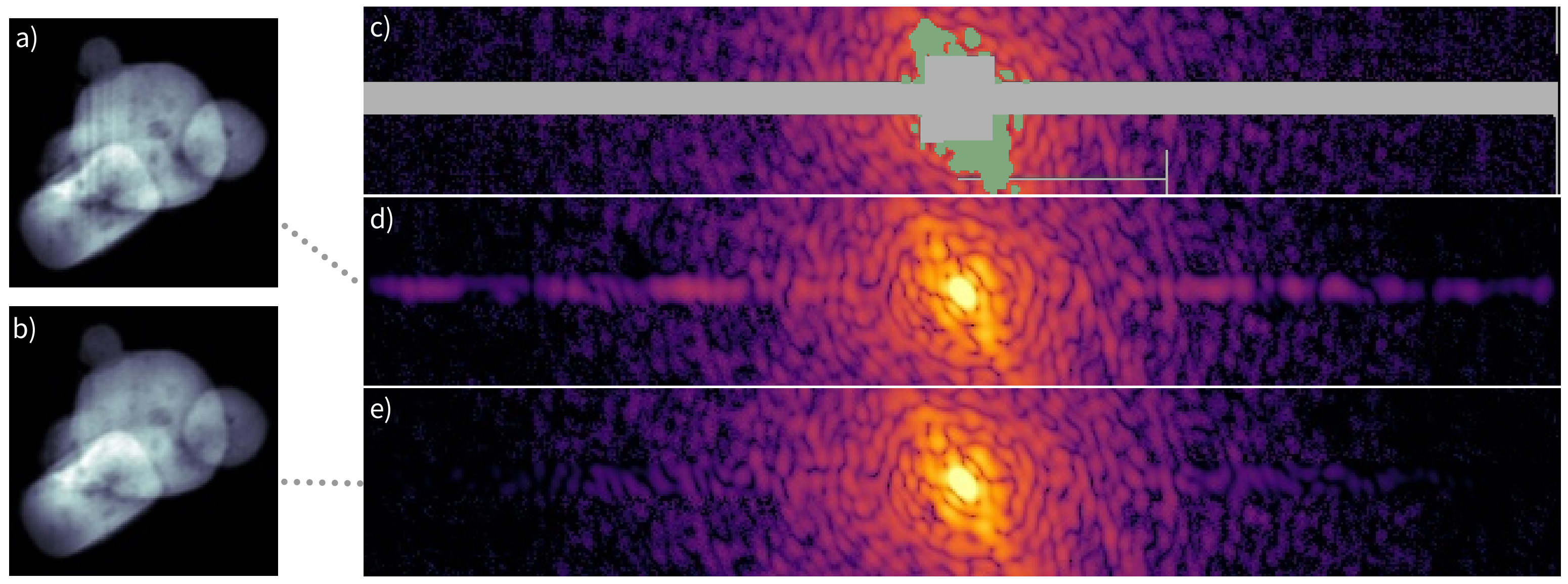}
  \caption{Effects of missing intensities in the detector's gap on the reconstructions. In c), a cutout of the experimental data, taken from Dataset B, is shown. The gray region identifies the missing values due to the detector's geometry, while the green one highlights the intensities missing due to pixels saturation. In a), the reconstruction is performed without constraints on the missing values shown in c). The reconstruction shows strong artifacts in the form of vertical lines in the reconstructed scattering density. Those artifacts come from the overestimation of the intensities in the detector's gap. This is highlighted in d), where the amplitude of the Fourier Transform of a) is shown. Subfigure b) shows the reconstruction performed by placing an upper bound to the missing intensities in the gap, as described in the main text, where the strong artifacts are now missing. The respective Fourier Transform in e) shows how, in this case, the overall amplitude of the intensities in the detector's gap are now fitting the overall intensity behavior.}
  \label{fig:bounds}
\end{figure*}

The problem of missing intensities due to the detector's tiles arrangement is particularly impacting the case of Dataset B (see Fig. \ref{fig:dataexamples}d), where the data has been acquired via the pnCCD detector of the SQS beamline at EuXFEL \cite{kuster20211}. The detector is composed of two modules, upper and lower, which can be independently moved. This arrangement is such that, in the horizontal direction, there is a region of unknown values at all $\vec{q}$ magnitudes, highlighted in gray color in Fig. \ref{fig:bounds}c. 
Such a large, spatially correlated, region of unknown values in Fourier space is challenging for reconstruction algorithms, which often tend to misestimate those values during the reconstruction process. This is well exemplified by the reconstruction shown in Fig. \ref{fig:bounds}a, where strong artifacts in the form of vertical lines arise from an overestimation of the reconstructed intensities in the horizontal gap (see Fig. \ref{fig:bounds}d).

A workaround to this problem, implemented in MPR, is to set upper bounds for those intensity values, based on the generic behavior of the scattering pattern. 
Those upper bounds can be calculated by considering the behavior of the scattering intensity as function of the distance from the detector's center. The scattering intensity $I(\vec{q})$ can be rewritten in terms of polar coordinates of the momentum transfer $\vec{q} = (q, \phi)$, i.e. $I(q, \phi)$. For a given magnitude of the momentum transfer $q$ (corresponding to a fixed scattering angle, i.e. to a fixed distance from the detector's center), the average scattering intensity $\mu_I (q)$ can be calculated as:
\begin{equation}\label{eq:avgint}
\mu_I (q)  = \left\langle I(q, \phi) \right\rangle_{\phi} = \frac{\int_0^{2\pi} d\phi \,  I(q, \phi) \Xi(q,\phi) }{\int_0^{2\pi} d\phi \, \Xi(q,\phi)}
\end{equation} 
where $\Xi(q,\phi)$ is the mask function described in Appendix \ref{subsec:init}, which assumes a value of $\Xi(q,\phi)=0$ where pixels are unknown, or $\Xi(q,\phi)=1$ otherwise. The function $\left\langle I(q, \phi) \right\rangle_{\phi}$ is commonly addressed as the \emph{radial profile} of the scattering. Still, Eq. \eqref{eq:avgint} does not take into account the fluctuations of the values that happen along the integration circumference. It is then necessary to take into account those fluctuations by calculating the standard deviation of the average intensity at a given magnitude of the momentum transfer, i.e.:
\begin{equation}\label{eq:stdint}
\sigma_I (q) =  \sqrt{\left\langle I^2(q, \phi) \right\rangle_{\phi} - \left\langle I(q, \phi) \right\rangle_{\phi}^2}
\end{equation}

The upper bound for the intensities $U(q)$, which is assumed then to be dependent only on the magnitude of $q$, is defined as:
\begin{equation}\label{eq:upperbound}
U(q) = \mu_I (q) + \eta \cdot \sigma_I (q)
\end{equation}
where $\eta > 0$ is a constant value. In practical terms, the upper bound for the missing intensity is set to be the average value of the known intensities at that scattering angle increased by $\eta$ times their standard deviation. The exact value for the parameter $\eta$ is a free parameter. In the context of this paper, this value is kept to $\eta=1.5$.

The upper bounds in Eq. \eqref{eq:upperbound} work upon the assumption that the missing intensities follow the same statistics as the known ones. This is a fair assumption when, for a given momentum $q$, the amount of known intensities is substantially larger than the unknown ones. As the gap between the detector's modules is fixed, while the length of the circumference where the integration is calculated increases with $q$, this condition is certainly met far away from the image center. Conversely, for low $q$ values close to the center, the relative amount of unknown intensities increases, up to the point where a statistical extrapolation from the known intensities becomes inadequate.
In the context of this paper, the upper bound in Eq. \eqref{eq:upperbound} is only used for those scattering angles where the relative amount of unknown intensities is within \SI{20}{\percent} of the total.
A further consideration is necessary for regions where the detector is saturated. First, the upper bound is not applied here. Second, if at a given magnitude of the momentum $q$ more than \SI{5}{\percent} of the pixels are saturated, then the upper bound $U_I$ is not applied.

Once the upper bound is defined based on Eq. \eqref{eq:upperbound} with the subsequent considerations, the next discussion point is to define how this upper bound is enforced during the reconstruction process. Such an operation is performed by a modification of the intensity projector, introduced in Eq. \eqref{eq:pm} in a simplified form. In fact, the projector $P_M$ replaces the current estimate of the Fourier amplitudes $\left| \tilde{\den}_{ij} \right|$ with the experimental ones, $M_{ij}=\sqrt{I_{ij}}$ only where those values are known, i.e. where $\Xi_{ij}>0$. The action of $P_M$, which creates a new density $\den^{\text{new}}$, can then be defined as:
\begin{equation}\label{eq:pm_full}
\tilde{\den}^{\text{new}}_{ij} = 
\begin{cases}
M_{ij} \,  \frac{\tilde{\den}_{ij}}{\left| \tilde{\den}_{ij}  \right|} & \text{if } \Xi_{ij}=1\\ 
\tilde{\den}_{ij} & \text{otherwise}
\end{cases} 
\end{equation}

The enforcement of the upper bounds for the unknown intensity values can be done via a modification of the intensity projector $P_M$, which now acts as:
\begin{equation}\label{eq:pm_bounds}
\tilde{\den}^{\text{new}}_{ij} = 
\begin{cases}
M_{ij} \,  \frac{\tilde{\den}_{ij}}{\left| \tilde{\den}_{ij}  \right|} & \text{if } \Xi_{ij}=1\\ 
U_{ij} \, \frac{\tilde{\den}_{ij}}{\left| \tilde{\den}_{ij}  \right|} & \text{if } \Xi_{ij}=0  \\ 
																		& \text{ and } U_{ij} \text{ is valid} \\
																		& \text{ and } \left| \tilde{\den}_{ij}  \right| > U_{ij}  \\ 
\tilde{\den}_{ij} & \text{otherwise}
\end{cases} 
\end{equation}
where $U_{ij}$ is the matrix that numerically encodes the upper bounds for each pixel coordinate calculated via Eq. \eqref{eq:upperbound}.

The improvements in terms of reconstruction quality derived from the use of the upper bounds for the unknown experimental data in the detector's gap are clearly observable in the reconstruction shown in Fig. \ref{fig:bounds}b (to be compared to the unconstrained version in Fig. \ref{fig:bounds}a). The corresponding Fourier intensities are shown in Fig. \ref{fig:bounds}e, where signs deriving from a misestimate of the unknown values are no longer visible (to be compared with the unconstrained result in Fig. \ref{fig:bounds}d).

It is worth noting that the presented method to mitigate this type of artifact is not strictly dependent on the MPR approach, and can be directly transferred to any imaging routine that makes use of standard iterative algorithms for phase retrieval. 

The constraint of upper bounds for missing data results in a clear improvement in the reconstruction quality for diffraction patterns that belong to Dataset B. In regards to Dataset A (see Fig. \ref{fig:dataexamples}a), the arrangement of the detector modules is such that, away from the central part, most of the intensities are recorded at all scattering directions. In this arrangement, the gap between the detector's components is not a main issue in terms of the quality of the reconstructions, which are naturally less prone to such artifacts.

Still, in both datasets, information at low momentum transfer $q$, i.e. close to the detector's center, is missing due to the presence of a hole or notch in the detector's modules (to allow for the transmission of the main FEL beam in order to avoid damage) and/or due to saturation of the detector's pixels (see the green region in Fig. \ref{fig:bounds}c). How the MPR algorithm deals with this aspect is discussed in the next section.

\subsection{Missing intensities at low scattering angles} \label{subsubsec:crossavg}

In general, the lack of diffraction data in the central part of the detector is a strong limiting factor for CDI. In fact, the central region of the detector corresponds to low momentum transfer (i.e. low scattering angles), which encode the low resolution information on the sample, and in particular:
\begin{enumerate}
\item The overall shape of the sample. It is harder to retrieve the correct support function, pivotal for a successful reconstruction and for the correct identification of the structural properties of the sample. 
\item The density profile at large length scales. This prevents iterative algorithms from correctly identifying the amplitude of the scattering density in real space. 
\end{enumerate}

This missing data has to be retrieved by iterative algorithms using the known intensities. This is theoretically possible because incorrectly retrieved intensities in the unknown central region of the detector cause an increase in density values outside the support function in real space (i.e., they are incompatible with the support constraint).

However, there are some specific intensity distributions in the central part that are still highly compatible with the support constraint. Those are known as ``weakly constrained modes'' \cite{thibault2006reconstruction} and can be added to a density $\rho^\text{sol}$, i.e. a solution to a specific phase retrieval problem, without significantly changing its error value. The possibility of constraining those modes becomes weaker and weaker with (i) larger samples (as the support function is wider and the problem less constrained) and (ii) larger region of missing intensities (as their effect on the density outside the support function is lower). For further details on the weakly constrained modes please refer to Ref. \cite{thibault2006reconstruction}.

During a reconstruction process, those \emph{unconstrained modes} tend to arise, especially due to the action of more \emph{ergodic} algorithms like HIO and RAAR. The crossover operation implemented in MPR (see Appendix \ref{subsec:crossover}) aims to augment the explorative power of standard IAs. As a side effect, the crossover operation further amplifies those unwanted modes. 

Within the MPR method implemented in SPRING, the occurrence of \emph{unconstrained modes} due to missing data in the central part of the detector can be mitigated by injecting information from the average reconstruction into the crossover operation. 
The average reconstructed intensities, calculated over the population $\left\{ \rec^p \right\}$, can be extracted in different ways. 
One possibility is to employ the amplitude of the Fourier Transform of the average reconstruction $\den^\text{avg}$, i.e.:

\begin{equation}\label{eq:avgint1}
\dot{M}^\text{avg}_{ij} = \left|\mathcal{F}\left[ \den^\text{avg} \right]_{ij} \right| = \left|\mathcal{F}\left[ \frac{\sum_p \rec^p \mem \den}{P}\right]_{ij} \right|
\end{equation}
The second possibility is to instead directly calculate the average of the retrieved Fourier moduli, i.e.
\begin{equation}\label{eq:avgint2}
\ddot{M}^\text{avg}_{ij} =  \frac{\sum_p  \left|\mathcal{F}\left[\rec^p \mem \den \right]_{ij} \right|}{P}
\end{equation}
It can be easily verified that $\dot{M}^\text{avg}_{ij} \leq \ddot{M}^\text{avg}_{ij}$, where the equal relationship only holds when all the densities  $\den^p$ have the same phase values in Fourier space. 
Due to the strongly differing phase values in particular in the first stages of the reconstruction process, $\dot{M}^\text{avg}_{ij}$ defined in Eq. \eqref{eq:avgint1} tends to be an underestimate of the correct Fourier moduli. On the other hand, $\ddot{M}^\text{avg}_{ij}$ tends to give an overestimate of the moduli, due to the strictly positive nature of the quantity $\left|\mathcal{F}\left[\den^p \right]_{ij} \right|$. These qualitative considerations lead us to the use of the average between the quantities in Eq. \eqref{eq:avgint1} and Eq. \eqref{eq:avgint2} as an estimate for the missing Fourier moduli. 

The information gained via Eq. \eqref{eq:avgint1} and Eq. \eqref{eq:avgint2} is then propagated into the population of reconstructions via the crossover operator (Appendix \ref{subsec:crossover}). There, the density values of three different \emph{parent} reconstructions, $\rec^a$, $\rec^b$ and $\rec^c$, are combined in reciprocal space following the operation described at line \ref{l:crossover:mix} of Alg. \ref{alg:crossover}. 
The parent reconstruction $\rec^a$ is then replaced with a modified reconstruction $\bar{\rec}^a$. The densities in Fourier representation $\rec^a \mem \tilde{\den}$ and $\bar{\rec}^a \mem \tilde{\den}$ have the same phase values. The modulus of $\bar{\rec}^a \mem \tilde{\den}$ is instead defined as:
\begin{equation}\label{eq:crossavg}
\left| \bar{\rec}^a \mem \tilde{\den}_{ij} \right| = C_a \cdot \frac{\dot{M}_{ij} + \ddot{M}_{ij}}{2} + (1-C_a) \cdot \left| \rec^a \mem \tilde{\den}_{ij} \right|
\end{equation}
The reconstruction $\bar{\rec}^a$ is then used in place of $\rec^a$ as a parent density for the crossover. 
In practical terms, Eq. \eqref{eq:crossavg} sets the amplitude of $\bar{\rec}^a \mem \tilde{\den}$ as a linear combination of the average moduli and the amplitude of the original reconstruction $\rec^a \mem \tilde{\den}$, depending on the parameter $C_a$. When $C_a=1$, $\bar{\rec}^a \mem \tilde{\den}_{ij}$ is obtained by completely replacing the moduli of $\rec^a \mem \tilde{\den}_{ij}$ with the average ones. 
When $C_a=0$, $\bar{\rec}^a \mem \tilde{\den}_{ij}$ is just a replica of the original density $\rec^a \mem \tilde{\den}_{ij}$.

\begin{figure*}
  \centering
  \includegraphics[width=1.\linewidth]{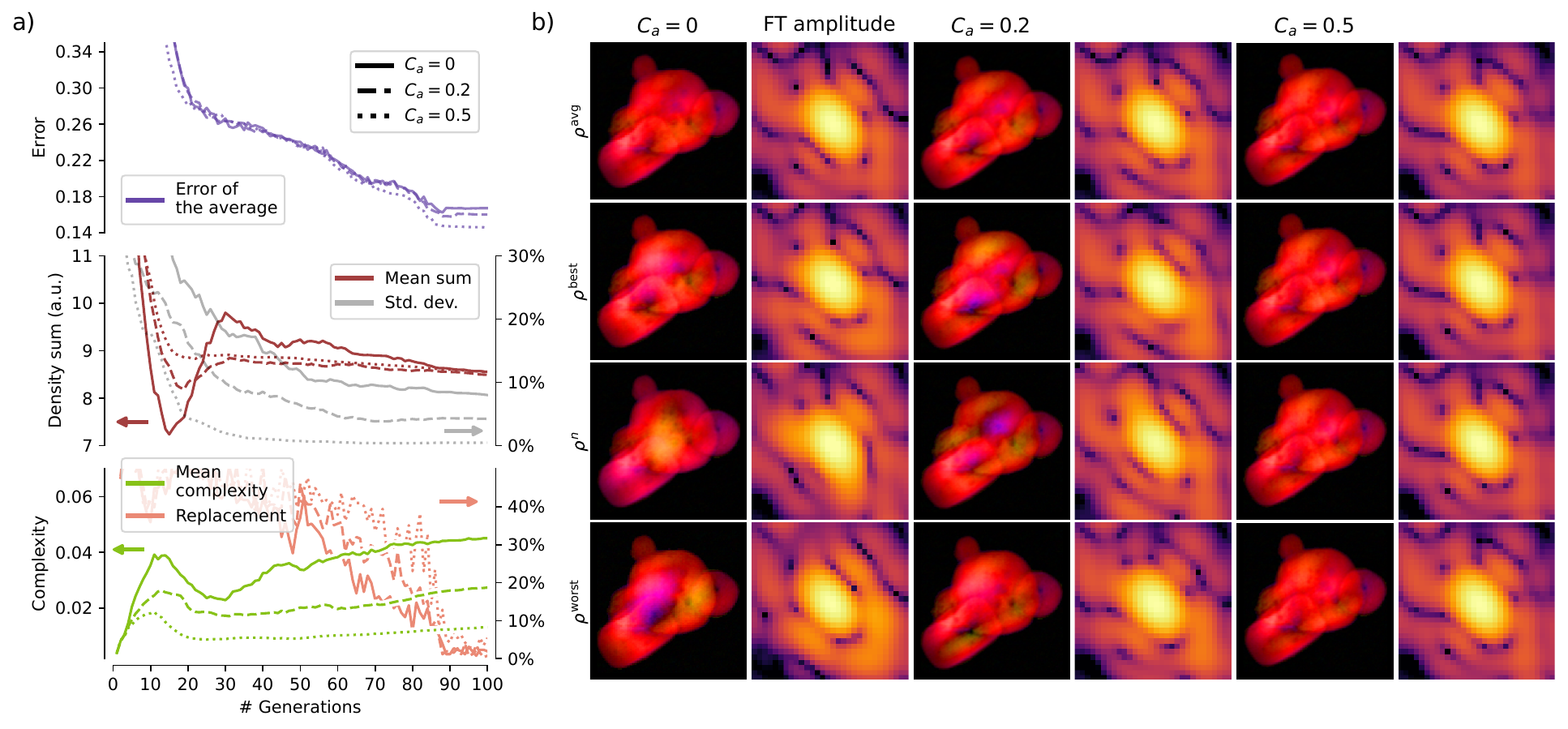}
  \caption{Effects of the presence of unconstrained modes on the reconstructions, performed on a diffraction image taken from Dataset B. In a), different performance indicators are reported as function of the algorithm's generations. Solid lines report the value of the indicators for the \emph{standard} crossover operation described in Appendix \ref{subsec:crossover}, which corresponds to $C_a=0$. Dashed and dotted lines describe instead the values for $C_a=0.2$ and $C_a=0.5$, respectively. The relative standard deviation (gray line) and the replacement factor (orange line) are expressed in $\%$, with the scale reported on the right axis of the respective plots. In b), insights into the status of population of reconstructions at the last generation of the algorithm (100) are reported for the three different values of $C_a$. In the first row, the average reconstruction is shown, indicated as $\den^\text{avg}$. The second row shows the reconstructions with the lowest error, $\den^\text{best}$. The third row reports the density in the middle of the errors distribution, $\den^n$. The last row depicts the density $\den^\text{worst}$ with the highest error. For each density, a cut-out of the modulus of its Fourier transform is also reported in logarithmic color scale, to highlight how the missing data in the central part of the detector affects the reconstructions.}
  \label{fig:complexity}
\end{figure*}

The behavior of SPRING for a spatially large sample, a particle agglomerate, is shown in Fig. \ref{fig:complexity} for three different values of the $C_a$ parameter.
Fig. \ref{fig:complexity}b is divided into three columns, corresponding to $C_a=0$, $C_a=0.2$ and $C_a=0.5$.  For each case, the average reconstruction is reported along with three representative individuals, similarly to what is reported in Fig. \ref{fig:monitor}. Here, the color scale encodes the phase component of the complex-valued reconstructed densities, following the color scheme in Fig. \ref{fig:init}. In the individual reconstructions, large-scale fluctuations of the density phases are recognizable, which come from the introduction of the unconstrained modes (apart from the Gaussian one, see Ref. \cite{thibault2006reconstruction}). 

It is possible to inspect the behavior of these unconstrained modes by tracking two quantities. First, the unconstrained modes affect the overall sum of the reconstructed density, i.e. $\left| \sum_{ij} \den_{ij} \right|$. This value corresponds to the central peak of the Fourier Transform and is theoretically encoded at the center of the diffraction pattern, i.e. $\left| \sum_{ij} \den_{ij} \right| = \sqrt{I(0,0)}$. As the central part of the pattern is missing, this value is not fixed and tends to fluctuate during the reconstruction process. It is well highlighted by tracking the mean value of the densities sum in the population of reconstructions, shown as a solid red line in Fig. \ref{fig:complexity}a for the case $C_a=0$ (no average information is injected into the crossover). The behavior shows that the density sum quickly drops in the first 15 generations, during the \emph{shape identification} phase, undershooting its final value (at generation $100$) by around \SI{20}{\percent}. In the following \emph{shape definition} stage, the density sum quickly recovers, now overshooting by  around \SI{10}{\percent} the final value, which is slowly reached during the \emph{density refinement} step. It is worth looking also at the behavior of the standard deviation of the density sum (expressed in $\%$ relative to the mean value), represented as a gray solid line in the same plot of Fig. \ref{fig:complexity}a, which slowly decreases towards a final value of around \SI{9}{\percent}. Thus, the width of the sums distribution within the population at the final generation $100$ is around the \SI{20}{\percent} of its value.

A second way of tracking the presence and the strength of unconstrained modes in the population is to define a \emph{complexity indicator} that estimates the fluctuations of the phase values in real space. For a density $\den$, the quantity $\Gamma$ can be defined as:
\begin{equation}\label{eq:complexity}
\Gamma = 1 - \frac{\left| \sum_{ij} \den_{ij} \right|}{\sum_{ij} \left| \den_{ij} \right|}
\end{equation}
The value of $\Gamma$ can assume values between $0$ and $1$. A value of $\Gamma=0$ implies that all the entries of the matrix $\den_{ij}$ have a constant phase value, i.e. there is no phase contrast in the reconstruction. As the refractive index of materials in the X-ray region of the electromagnetic spectrum is typically very close to $1$, no (or very small) phase shifts are expected (there are conditions where this approximation is not valid, as discussed in Sec. \ref{subsec:complex}). Instead, the higher the value of $\Gamma$, the stronger are the density phases fluctuations in real space. 
The behavior of the \emph{complexity} indicator $\Gamma$ for the case $C_a=0$ is revealed in Fig. \ref{fig:complexity}a as a green solid line, which reports the mean value within the population of reconstructions. The value of $\Gamma$ quickly raises from $0$ (as densities are initialized as fully real-valued, see Table \ref{tab:params} and Appendix \ref{subsec:init}) up to $0.04$ and then stabilizes right above $0.02$ in the first two stages of the reconstruction. During the following \emph{shape refinement} stage, its value starts to increase again in an apparently divergent behavior, reaching its maximum value in the last generations. 

The high values of $\Gamma$ within the populations are a direct implication of the fluctuations in the density values visible in the first column of Fig. \ref{fig:complexity}b, especially in rows 2-4, where three individual reconstructions of the population $\left\{ \rec^p \right\}$ are shown. Those fluctuations, whose shape and strength widely differ among the single individuals, can be related to the variations of the reconstructed missing intensities in the central part of the diffraction pattern, shown on the right side. First of all, the distribution of the reconstructed intensities varies strongly among the reconstructions. Second, they present strong asymmetries with respect to the center of the diffraction image, which is the sign of a remarkable deviation from a real-valued density distribution in real space. 

The impact of different values of the $C_a>0$, i.e. $C_a=0.2$ and $C_a=0.5$, on the reconstruction process are also reported in Fig. \ref{fig:complexity}a. For what concerns the mean value of the density sum (red line in \ref{fig:complexity}a) the large fluctuations of the value described before in the case of $C_a=0$ are increasingly dampened for $C_a=0.2$ (dashed red line) and $C_a=0.5$ (dotted red line). In particular, the sum value quickly stabilizes within the first 15 generations for $C_a=0.5$, with only slight adjustments towards its final value over the remaining iterations. The stronger ``constraint'' on the missing intensities is also directly visible in the relative standard deviation within the reconstructions populations of the sum values (gray line in \ref{fig:complexity}a), which drops much quicker with the proceeding of the iterations, reaching a final value below \SI{5}{\percent} in the case of $C_a=0.2$ and close to \SI{0}{\percent} for $C_a=0.5$. It is worth noting here that, despite the substantially different behavior of the sum, the final value reached at generation $100$ is almost identical for the three cases. 

Similar considerations apply for the complexity indicator, reported as a green line in Fig. \ref{fig:complexity}a. Its average value within the population is increasingly lower for higher values of $C_a$. This feature is directly visible by observing the individual reconstructions in Fig. \ref{fig:complexity}b. In particular, the large-scale fluctuations of the phase values in real space (and the corresponding asymmetry in the central cutout of the Fourier amplitudes) are weaker, but still recognizable, for $C_a=0.2$ (see third and fourth columns of Fig. \ref{fig:complexity}b). It is instead hard to spot differences, both in real space and in Fourier amplitudes, in the case of $C_a=0.5$. 

Despite the large fluctuations observed for $C_a=0$ (and still present in $C_a=0.2$) the final average density presents very similar features in all the three cases (first row of \ref{fig:complexity}b), both in terms of phase fluctuations in real space and in terms of retrieved missing amplitudes in Fourier space. This observation hints towards the fact that, even in the case of $C_a=0$, despite the prominent arising of the unconstrained modes in the individual reconstructions, those modes are randomly distributed and tend to cancel out in the averaging operation.

However, the use of the average Fourier modulus via the $C_a$ parameter unveils its importance by comparing the \emph{replacement factor} of the crossover operation (see Sec. \ref{subsec:monitor}), shown as orange line in Fig. \ref{fig:complexity}a. During the \emph{density refining} stage (and in particular from generation $50$), the replacement indicator is systematically higher for higher values of $C_a$, meaning that the crossover operator is more effective in feeding the population with new reconstructions that perform better than the previous generation. The direct implication is that the error of the average reconstruction is lower, as reported by the upper plot in Fig. \ref{fig:complexity}a, i.e. the error value is better optimized.

\section{Computational cost and performance} \label{subsec:timing}

\begin{figure*}
  \centering
  \includegraphics[width=\linewidth]{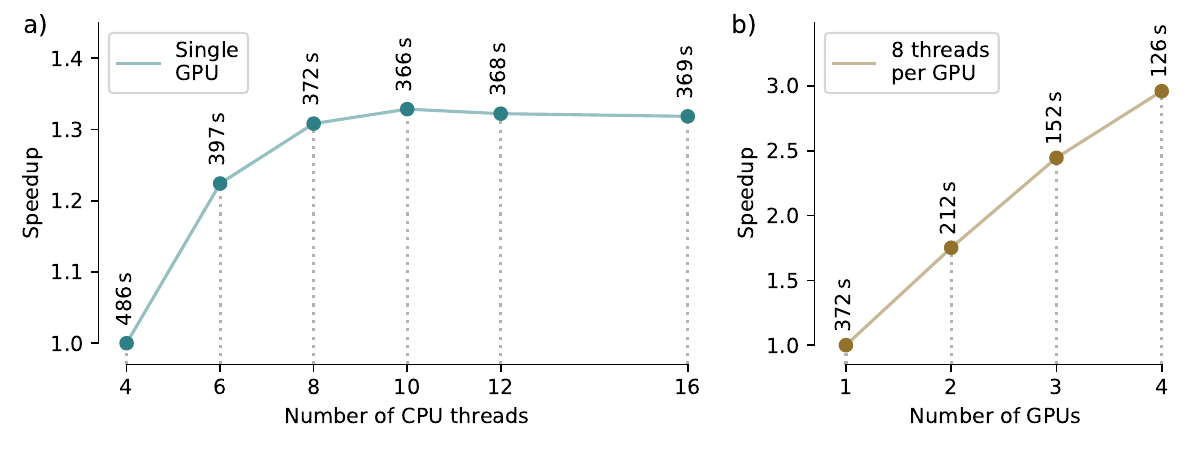}
  \caption{Time to solution and speedup for a full reconstruction process with MPR, with different combinations of CPU threads and GPUs in use. In a), a single GPU is employed with increasing number of CPU threads. The speedup is relative to the case where the number of CPU threads is set to 4. The time to solution is reported in seconds above the markers. In b), The speedup is reported as function of the number of GPUs in use, with a constant number of CPU threads per GPU. The speedup value is relative to the case where only one GPU is employed. }
  \label{fig:timing}
\end{figure*}

The parallel execution of several individual reconstructions is a key ingredient of the MPR approach and it comes
at a computational cost. In this section we briefly discuss this aspect and give a reference for the time to solution that a user may expect. The presented timing tests, summarized by Fig. \ref{fig:timing}, are performed with the MPR implementation provided by the \emph{spring} python module, which can exploit multiple CPU cores as well as multiple Graphical Processing Units (GPUs) in a shared-memory environment. The part of the software executed on the GPU is written in CUDA\textsuperscript{\textregistered} programming language  \cite{CUDA}, which is designed only for GPUs produced by NVIDIA\textsuperscript{\textregistered}.

Among the three operations cyclically performed by MPR, the computational cost of the \emph{crossover} and \emph{selection} is, at a first approximation, negligible compared to the computational effort required in the \emph{self-improvement}. There, in fact, thousands of FT have to be executed for each reconstruction in the populations $\left\{ \rec^p \right\}$ and  $\left\{ \rec^p_\text{new} \right\}$.

In theory, the time to solution of a full reconstruction process with MPR is linearly dependent on the number of iterations of iterative phase retrieval algorithms ($\iter$ and $\iter_\text{ER}$), the size of the population $P$ and the number of generations $G$ (see Table \ref{tab:params}). It is furthermore higher than linearly dependent on the number of entries in the matrix containing the diffraction data, $N_p^2$, which is at least $N_p^2 \log(N_p^2)$, thanks to the properties of the Fast Fourier Transform employed for the FT calculation \cite{cooley1965algorithm}.

Timing and scaling tests for different combinations of number of CPU threads and GPUs are reported in Fig. \ref{fig:timing}. These tests are performed on a server equipped with two AMD Epyc\textsuperscript{\tiny TM} 7302 CPUs with 16 cores each, for a total of 32 physical computing cores. The system is accelerated by four NVIDIA GeForce RTX\textsuperscript{\tiny TM} 3090 high-end consumer-level GPUs.
The values of the MPR parameters that influence the time to solution have been set to their default values (see Table \ref{tab:params}), i.e. $\iter=\iter_\text{ER} = 40$, $P=128$, $G=100$ and $N_p=512$.

Fig. \ref{fig:timing}a reports the timing and scaling performance with a single GPU in use and increasing number of CPU threads. We decided to use 4 threads as a starting value, as this is typically the minimum number of CPU cores provided nowadays even on a lower-end consumer-level computer accelerated by a GPU. As most of the calculations are off-loaded to the GPU, a strong improvement in performance should not be expected with the increasing number of CPU cores in use. Still, NVIDIA GPUs can handle communications with CPU threads in parallel, overlapping communication and computation up to some degree. In this specific hardware configuration, an appreciable gain in performance is visible for up to 8 CPU threads. 
Fig. \ref{fig:timing}b shows, instead, the timing and scaling of a single reconstruction process for increasing number of GPUs, from 1 to 4. The number of CPU threads per GPU is kept constant and equal to 8.

The tests reported in Fig. \ref{fig:timing} are only a rough estimate of the performance that can be expected for a given hardware configuration. They are anyway enough to conclude that, depending on the number and performance of the available GPUs, the time to solution can vary from a few to some minutes. The use of GPU accelerators designed for data centers, as well as the development expected in the upcoming years in terms of GPU performance, boosted by Artificial Intelligence, will allow a sub-minute time to solution for this MPR configuration. 

Relevant speedups in the time to solution can be furthermore obtained by tuning the binning and rescaling of the experimental data. All the tests in this manuscript have been performed by rescaling the data via binning operation to a size of $512 \times 512$ (i.e. $N_p=512$), which ensures a sufficient degree of \emph{oversampling} to handle any practically tractable diffraction pattern. Still, in many cases, the sample is sufficiently small to rescale the diffraction data to $N_p=256$, thus providing at least a four-fold speedup in the time to solution of MPR, reaching the sub-minute time scale.


\bibliographystyle{ieeetr}
\bibliography{bibliography}

\end{document}